\documentclass{sig-alternate-10pt}
\usepackage{times}

\usepackage[hyphens]{url}

\usepackage{graphics}
\usepackage{color}
\usepackage{multirow}
\usepackage{subfigure}
\usepackage{graphicx}
\usepackage{pbox}
\usepackage{comment}
\usepackage{mdwlist,enumitem}
\usepackage{epstopdf}

\usepackage{color, colortbl}
\definecolor{Gray}{gray}{0.93}

\usepackage{pifont}

\usepackage{xcolor}
\definecolor{Orange}{rgb}{1,0.5,0}
\definecolor{blue}{rgb}{0,0,1}

\begin{document}

\clubpenalty=10000
\widowpenalty=10000


\setlength{\paperheight}{11in}
\setlength{\paperwidth}{8.5in}
\setlength{\pdfpageheight}{\paperheight}
\setlength{\pdfpagewidth}{\paperwidth}


\title{Anatomy of the Third-Party Web Tracking Ecosystem}


\numberofauthors{1} 
%
\author{
%
\alignauthor
Marjan Falahrastegar$\dag$, Hamed Haddadi$\dag$ $\ddag$, Steve Uhlig$\dag$, Richard Mortier$\diamond$\\
       \affaddr{Queen Mary University of London$\dag$, Qatar Computing Research Institute$\ddag$, University of Nottingham$\diamond$} \\
}
\maketitle

\newpage

\begin{abstract}
The presence of third-party tracking on websites has become customary. However, our understanding of the third-party ecosystem
is still very rudimentary. We examine third-party trackers from a geographical perspective, observing the third-party tracking 
ecosystem from 29 countries across the globe. When examining the data by region (North America, South America, Europe, East Asia, 
Middle East, and Oceania), we observe significant geographical variation between regions and countries within regions. We find 
trackers that focus on specific regions and countries, and some that are hosted in countries outside their expected target tracking 
domain. Given the differences in regulatory regimes between jurisdictions, we believe this analysis sheds light on the geographical 
properties of this ecosystem and on the problems that these may pose to our ability to track and manage the different data silos 
that now store personal data about us all.
\end{abstract}

\keywords{Advertisement, Privacy, Analytics, Trackers}

\category{K.4}{COMPUTERS AND SOCIETY}{Privacy}

\section{Introduction}
\label{sec:intro}

The rise in use of personal data and sophisticated algorithms on individuals' online browsing behaviour and interests has lead to increasing presence of third-party advertising and analytics services on the Internet and the mobile web~\cite{vallina2012breaking,Krishnamurthy,marjanTMA}. These services aim to build a \emph{user profile} by collection, aggregation, and correlation of an individual's browsing behaviour, demographics, interests, and temporal/spatial patterns of behaviour (e.g., through smartphone localisation, or location check-ins on Online Social Networks). While these services are vital for the online economy, there are complex debates over privacy issues which are caused directly or indirectly (e.g.,~misusing ad tracker cookies to identify individuals~\cite{nsa}) by such services. Despite legal or regulatory efforts on personal data collection and storage, existence of thousands of these third party services~\cite{marjanTMA} across the world under different names and legal constraints makes holding individual businesses responsible for their actions a challenging task.

In this paper we investigate the geographic diversity and footprint of thousands of third party domains across the world, using measurements from tens of vantage points distributed across nearly all continents. 
As a result, we shed light on the third party tracking and analytics industry across legal and geographic boundaries. Several recent works have highlighted the footprint of companies such as Google as dominant corporations and United States as a specific prevalent country in the third-party tracking market~\cite{castellucia:hal-00832784,Krishnamurthy}. Considering the role of technology, legislation and economics in privacy problem~\cite{Krishnamurthy:summer:2010}, we believe that understanding regional differences in the tracking ecosystem is essential for providing effective privacy regulations, in addition to ability to understand the trade of personal data.

We gathered data from 28 countries using PlanetLab\footnote{\url{http://www.planet-lab.org/}} nodes (\S\ref{sec:data}). We first analyse our dataset by dividing our data on a regional basis (\S\ref{sec:regions}), considering the major and minor players in each region separately. We find a significant difference in the presence and dominance of local third-parties in different regions. While we observe an even distribution of local third-parties in countries within Europe, East Asia, Oceania and South America, in contrast, Turkey and Israel in the Middle East appear much stronger in terms of their number of local third-parties. 
We then analyse our data using the country code TLD (Top Level Domain name) of each tracker (\S\ref{sec:countries}) to reveal a complex, interwoven set of cross-country relationships between third-party tracking services. We find extensive presence of European third-party trackers in the popular websites of East Asia and the Middle East. In particular, Germany and Russia third-parties are present across popular websites of all investigated countries in our dataset. Similarly, third-parties based in North America, mostly US, are broadly embedded in popular websites of the Middle East. We hypothesise that the reasons for the observed cross-country tracking is related to the substantial differences in privacy regulation in these countries. After a discussion of related work (\S\ref{sec:related}), we conclude (\S\ref{sec:conclusion}).
\section{Data Collection}
\label{sec:data}

\begin{table}
  \small
  \centering
  \renewcommand{\arraystretch}{1.2}
  \begin{tabular}{ l p{0.65\columnwidth} }
    \textbf{Region} & \textbf{Country}\\ \hline
    North America & Canada, US \\\hline
    South America & Argentina, Brazil, Ecuador\\ \hline
    Europe        & Belgium, France, Germany, Greece, Hungary, Italy, Netherlands, Norway, Russia, Slovenia, Sweden, United Kingdom\\ \hline
    East Asia     & China, Hong Kong, Japan, Korea, Taiwan\\\hline
    Middle East   & Israel, Jordan, Qatar, Turkey\\\hline
    Oceania       & Australia, New Zealand  \\\hline
  \end{tabular}
  \caption{\label{tab:countries}The countries for which we collected data and their assigned region.}
\end{table}

In this section we describe our data collection methodology. We extended Krishnamurthy's Firefox extension~\cite{Krishnamurthy:footprint} to log cookies, Etags\footnote{\url{http://www.w3.org/Protocols/rfc2616/rfc2616-sec14.html#sec14.19}}, and browser local storage. We then ran this script against the Alexa top-500 popular websites in each country listed in Table~\ref{tab:countries}, storing details of the observed third-party for later analysis. All our data was obtained between 28 March 2014 and 28 April 2014. To minimise pollution between consecutive visits, a bash script creates a new user profile and ensures that the local cache is cleared before each website is visited and the extension runs. We used PlanetLab's infrastructure~\cite{planetlab} nodes in 28 different countries to gain access across the globe. In addition to the PlanetLab servers, we also ran our scripts on a computer located in Qatar as PlanetLab's coverage in the middle-east is not as strong as elsewhere. Unfortunately, a paucity of PlanetLab nodes in Africa coupled with the failure of our scripts to complete successfully on the few nodes in Africa that we could try, we cannot present data pertaining to Africa.

To identify third-party websites, our Firefox extension employs the combination of the \emph{domain} and \emph{adns} approaches explained by Krishnamurthy \& Wills~\cite{Krishnamurthy}. A third-party site is identified as one whose second-level domain and ADNS (Authoritative DNS) server differ from the second-level domain name and authoritative DNS server of the origin site. Use of the authoritative DNS server allows us to classify cases such as \url{bbc.co.uk} and \url{bbci.co.uk} correctly, observing that both belong to the same company even though the second-level domains are different.

In visiting the Alexa top-500 websites in 28 countries from different regions of the world, we visited a total of 6497 unique websites and identified 6817 third-party trackers. We observed the presence of third-parties on the over 80\% of the visited websites. Qatar (814), Korea (769) and Hong Kong (726) are the top three countries in terms of number of third-party trackers, while the United Kingdom (397), Jordan (330) and Belgium (274) are the bottom three. We group countries into six geographical \emph{regions}: North America, South America, Europe, East Asia, Oceania and the Middle East. Table~\ref{tab:countries} shows the investigated countries and the regions to which they belong. Overall we detect the highest numbers of third-party trackers in Europe (3378) and East Asia (2009). Normalising by the number of countries in each region, we see that the North America, Oceania and the Middle East are the regions with the highest average numbers of third-party services.


\section{Regional Analysis}
\label{sec:regions}


\begin{figure}
  \includegraphics[trim=0 0 0 50,clip=true,width=.50\textwidth]{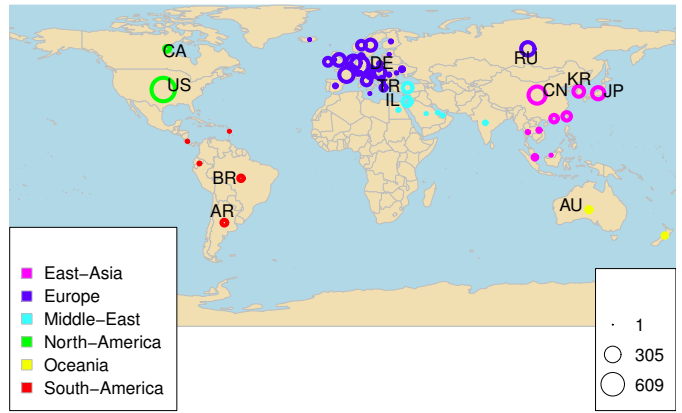}
  \caption{The strength of countries in terms of number of local third-party services.}
  \label{fig:geo}
\end{figure}

\begin{figure}
  \includegraphics[width=.50\textwidth]{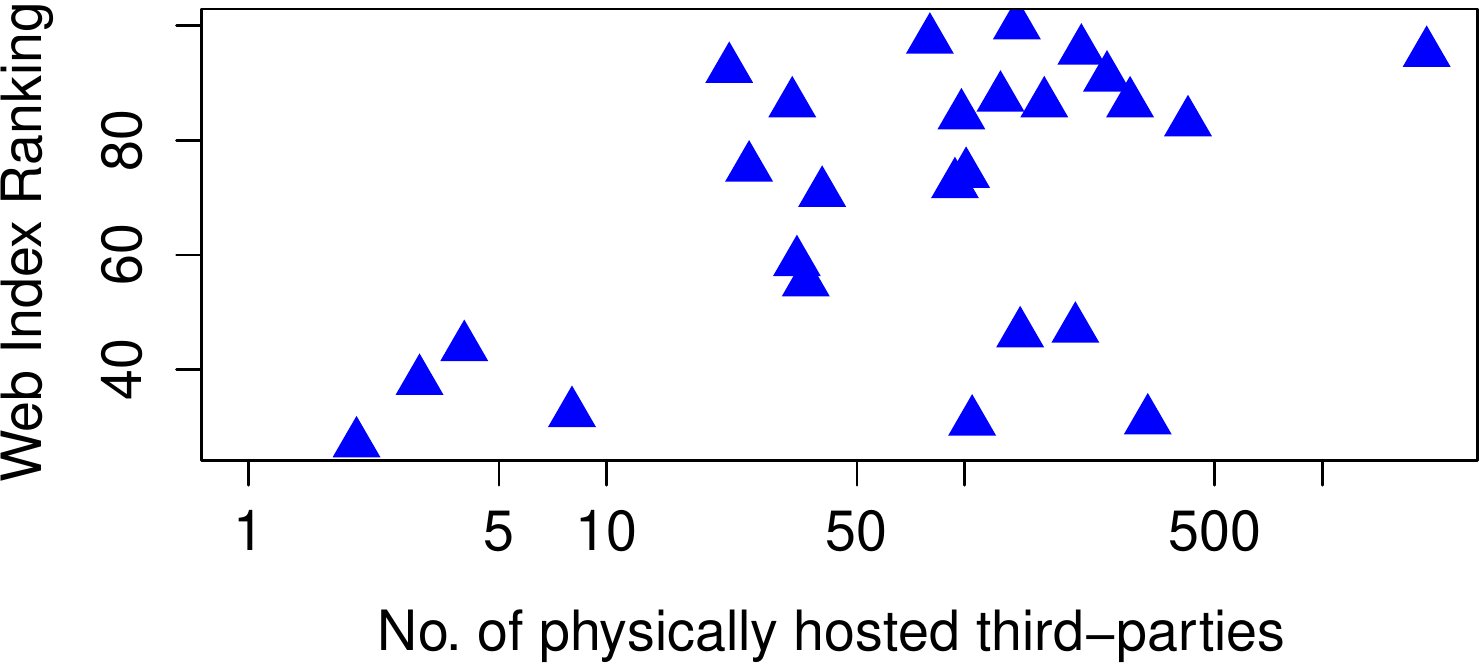}
  \caption{\label{fig:scatter} Web Index ranking vs Locally hosted third-parties}
\end{figure}

We begin our regional analysis by counting the number of third-parties which are physically hosted in the countries of our study. We rely on the \url{geoiplookup} utility to determine the country in which each observer third-party resides. Our results are shown in Figure~\ref{fig:geo}, with circle diameter representing the number of third-parties located in each country. We can see that locally hosted third-parties in countries across Europe, East Asia and Oceania regions are relatively evenly distributed, whereas, in North and South America and the Middle East there are substantial variations. For example, in the Middle East, Turkey and Israel have many more local third-parties than other countries in that region. In general we found North America, Germany and China with the highest number of locally hosted third-parties. After looking at this result one natural question that may arise is the correlation between Web-based advancement and number of locally hosted third-parties across countries. To answer this question, we used the Web Index~\cite{webindex} that is provided by WWW Foundation led by Tim Berners-Lee. The index , first released in 2012 and updated in 2013, measures the contribution of Web in 81 countries using four factors: "Universal Openness" for communication infrastructure, "Freedom and Openness" for citizen rights of information, opinion and online privacy, "Relevant content" for accessibility of relevant information based on gender and language, "Empowerment" for impact of the Web on society, economy and politics. Figure ~\ref{fig:scatter} presents the scatterplot for Web Index ranking against locally hosted third-parties per country. We observe that the majority of countries with high ranking have actually high number of locally hosted third-parties. Turkey, Hungary, Russia and China constitute four exceptions with over 100 locally services while they are ranked below 50. 


\begin{figure}
  \includegraphics[width=.45\textwidth]{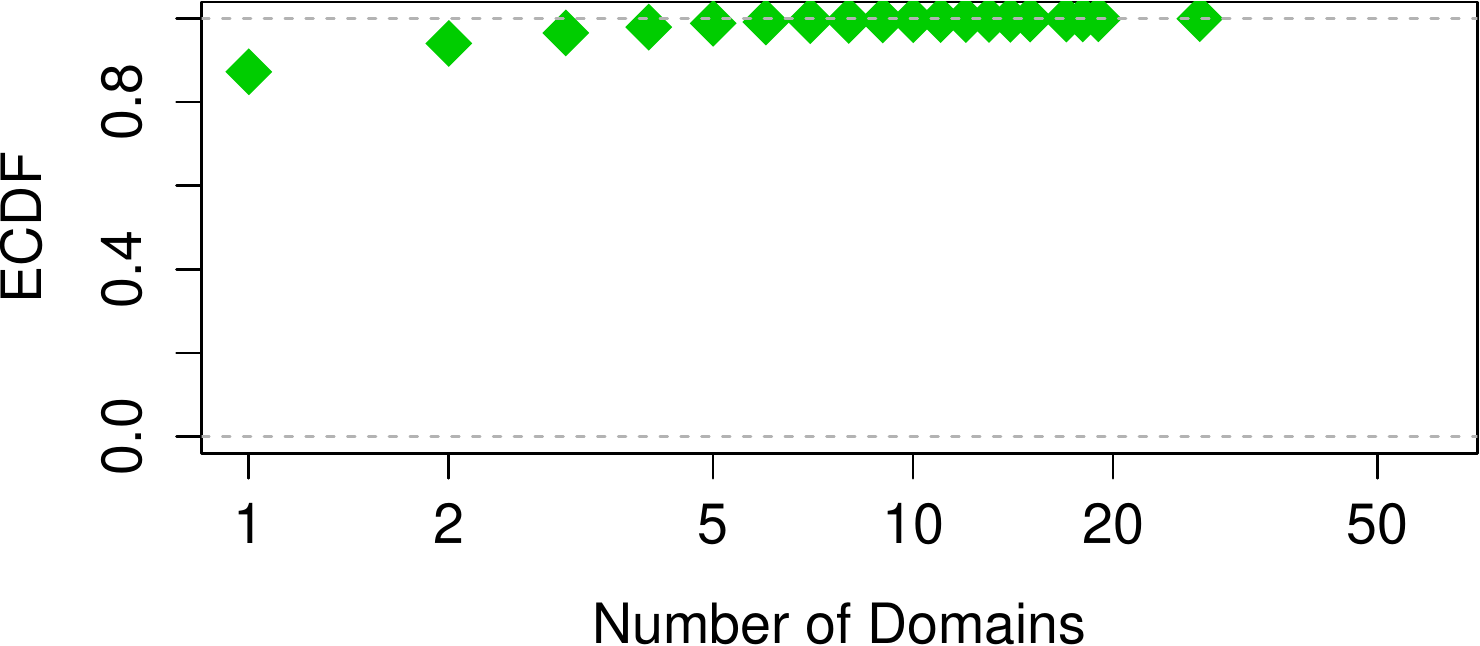}
  \caption{\label{fig:ecdf}Distribution of aggregated domains presents the existence of companies with tens of third-party trackers.}
\end{figure}

\begin{table}
  \small
  \centering
  \renewcommand{\arraystretch}{1.2}
  \begin{tabular}{ l l }
    \textbf{Company(\#Domains)} & \textbf{Company(\#Domains)}\\ \hline
google\_family (42)	&	elender.hu (9) \\ \hline
verisign\-grs.com (27)	&	gabia.com (9) \\ \hline
microsoft.com (19)	&	indom.com (9) \\ \hline
aol\_family (18)	&	schibsted-it.no (9) \\ \hline
sakura.ad.jp (17)	&	taobao.com (9) \\ \hline
first\-ns.de (15)	&	t\-online.hu (9) \\ \hline
netnames.net (15)	&	iponweb.net (8) \\ \hline
transip.nl (15)	&	knet.cn (8) \\ \hline
yahoo\_family (14)	&	sanomaonline.hu (8) \\ \hline
register.it (13)	&	baidu\_family (7) \\ \hline
sina\_family (12)	&	adresseavisen.no (7) \\ \hline
conversant\_family (11)	&	atcom.gr (7) \\ \hline
qq.com       (11)	&	bbend.com (7) \\ \hline
registercom (10)	&	comodogroup.com (7) \\ \hline
regtime.net (10) 	&	ebay.com (7) \\ \hline
  \end{tabular}
  \caption{\label{tab:co_agg}Top-30 identified companies with the highest number of aggregated domains.}
\end{table}

\begin{table}
  \small
  \centering
  \renewcommand{\arraystretch}{1.2}
  \begin{tabular}{ l p{0.65\columnwidth} }
    \textbf{Company} & \textbf{Domain}\\
    \hline
    AddThis
    & addthis.com, addthiscdn.com, addthisedge.com \\ \hline

    AOL
    & aol.com, advertising.aol.com, atwola.com, advertising.com, adsonar.com,
    Tacoda.com, pictela.net, huffingtonpost.com, huffpost.com, huffpo.net,
    mapquestapi.com, 5min.com, aolcdn.com, {goviral-content.com}, srvntrk.com
    blogsmithmedia.com, mirabilis.com, mqcdn.com \\ \hline

    Adobe
    & omniture.com, 2o7.net, demdex.net \\ \hline

    Amazon
    & amazonaws.com, images-amazon.com, cloudfront.net \\ \hline

    AudienceScience
    & revsci.net, wunderloop.com \\ \hline

    Baidu
    & baidu.com, baidustatic.com, hao123.com, hao123img.com, bdstatic.com,
    bdimg.com, hao123.com.br \\ \hline

    Burst Media
    & burstnet.com, blinkx.com \\ \hline

    ComScor
    & voicefive.com, scorecardresearch.com, securestudies.com, sitestat.com
    \\ \hline

    Conversant (ValueClick)
    & conversantmedia.com, awltovhc.com, kdukvh.com, qksrv.net, apmebf.com,
    ftjcfx.com, tqlkg.com, yceml.net, dotomi.com, mediaplex.com, lduhtrp.net
    \\ \hline

    Facebook
    & facebook.com, facebook.net, fbcdn.net \\ \hline

    Google
    & doubleclick.net, youtube.com, blogblog.com, android.com, 	
      ajax.googleapis.com,    {googlesyndication.com}, 
      doubleclick.com, {youtube.googleapis.com}, blogger.com,
      {channelintelligence.com}, {content.googleapis.com},
      {googletagmanager.com},  2mdn.net, {youtube-nocookie.com}, {blogger-comments.googlecode.com}, eedburner.com, {fonts.googleapis.com},
      {googleusercontent.com}, ytimg.com, , blogspot.com, gmodules.com,
      goo.gl, googlevideo.com, {wordtechnews.blogspot.com}, invitemedia.com,
      {googleadservices.com}, gstatic.cn, 
      ggpht.com, orkut.com, googleadsserving.cn, gstatic.com, 
      recaptcha.net, {google-analytics.com}, {javaplugins.googlecode.com}, 
      urchin.com, googleapis.com, {maps.googleapis.com}, 
      googlecode.com, {translate.googleapis.com}, 
      google.com, {www.googleapis.com}, 
      {googlecommerce.com}  \\\hline

    Nielson
    & mrworldwide.com, nielson.com\\ \hline

    Quantcast
    & quantcast.com, quantserve.com \\ \hline

    RadiumOne
    & radiumone.com, gwallet.com, po.st \\ \hline

    247 Real Media
    & realmediadigital.com, realmedia.com, rmlacdn.net, 247realmedia.co.kr
    \\ \hline

    Sina
    & sinajs.cn, sinaimg.cn, leju.com, weibo.com, sinauda.com, sinajs.js,
    wcdn.cn, sinahk.net, sina.com.cn, sinacdn.com, appsina.com,  sinahk.net          \\ \hline
    
    Sizmek
    & serving-sys.com, peer39.net, republicproject.com \\ \hline

    Twitter
    & twitter.com, ,twimg.com \\ \hline

    Yahoo
    &  yahoo.com, flickr.com, yieldmanager.com, bluelithium.com	,	overture.com, yahooapis.com, staticflickr.com, yldmgrimg.net,
      maktoob.com, xtendmedia.com, yahoo.net, sstatic.net,
     yimg.com, zenfs.com	\\\hline
  \end{tabular}
  \caption{\label{tab:family}Top-20 companies and their third-party domains.}
\end{table}

We carry on our analysis by identifying dominant third-parties in each region  after aggregating third-parties within their parent company, identified through a combination of three methods. First, we used Collusion's dataset~\cite{collusion} to detect third-parties belonging to the same company. We manually inspected this dataset for any changes using websites and wiki pages of the companies involved. Second, we used the e-mail addresses of third-party domains obtained by querying their SOA (Start of Authority)record (i.e.,~reapplying the adns method). However, the email address is often unhelpful if it is a general account from a cloud, CDN or DNS service. For example, \url{awsdns-hostmaster@amazon.com} is the email address of all third-parties hosted on Amazon Web Services, and \url{dns-admin@google.com} is assigned for all services hosted on Google App Engine. We identified the unhelpful email addresses if their email domain name belongs to the known CDN and DNS services, or keywords in the email domain indicate such services. Intead for these cases, we used the organization indicated in their \url{whois} records when available, else we assumed the third-party has no parent company. 

The distribution of aggregations we carried out is shown in figure~\ref{fig:ecdf}. The size of the parent companies varies considerably: some appear to own tens of  third-party trackers while others have fewer than five. Table~\ref{tab:co_agg} shows the top-30 identified companies with the highest number of aggregated third-parties. The well-known advertising-related (e.g.,~analytics, ad trackers) companies are presented in \textit{company name\_family} format. Moreover, the aggregated domains of such companies are listed in Table~\ref{tab:family}. We found, unsurprisingly, that Google, AOL and Yahoo appear to own the largest number of third-party trackers. We present the hierarchy of the top-four big companies, acquisitions and their trackers in Appendix~\ref{App:appendix}.




\begin{figure*}
  \centering

  \subfigure[North-America]{%
    \includegraphics[width=0.49\textwidth]
                    {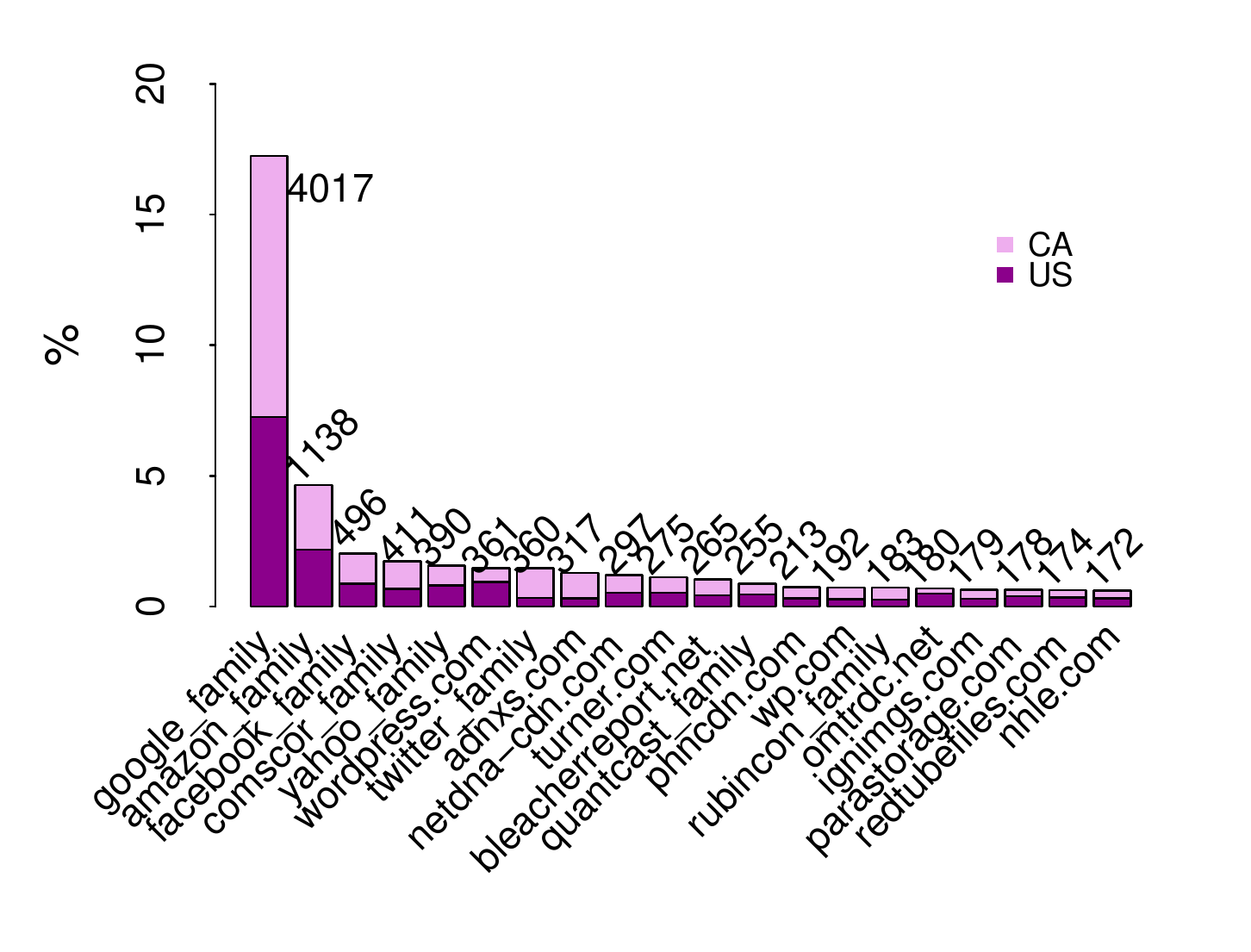}
                    \label{fig:topus}
  }
  \subfigure[South-America]{%
    \includegraphics[width=0.49\textwidth]
                    {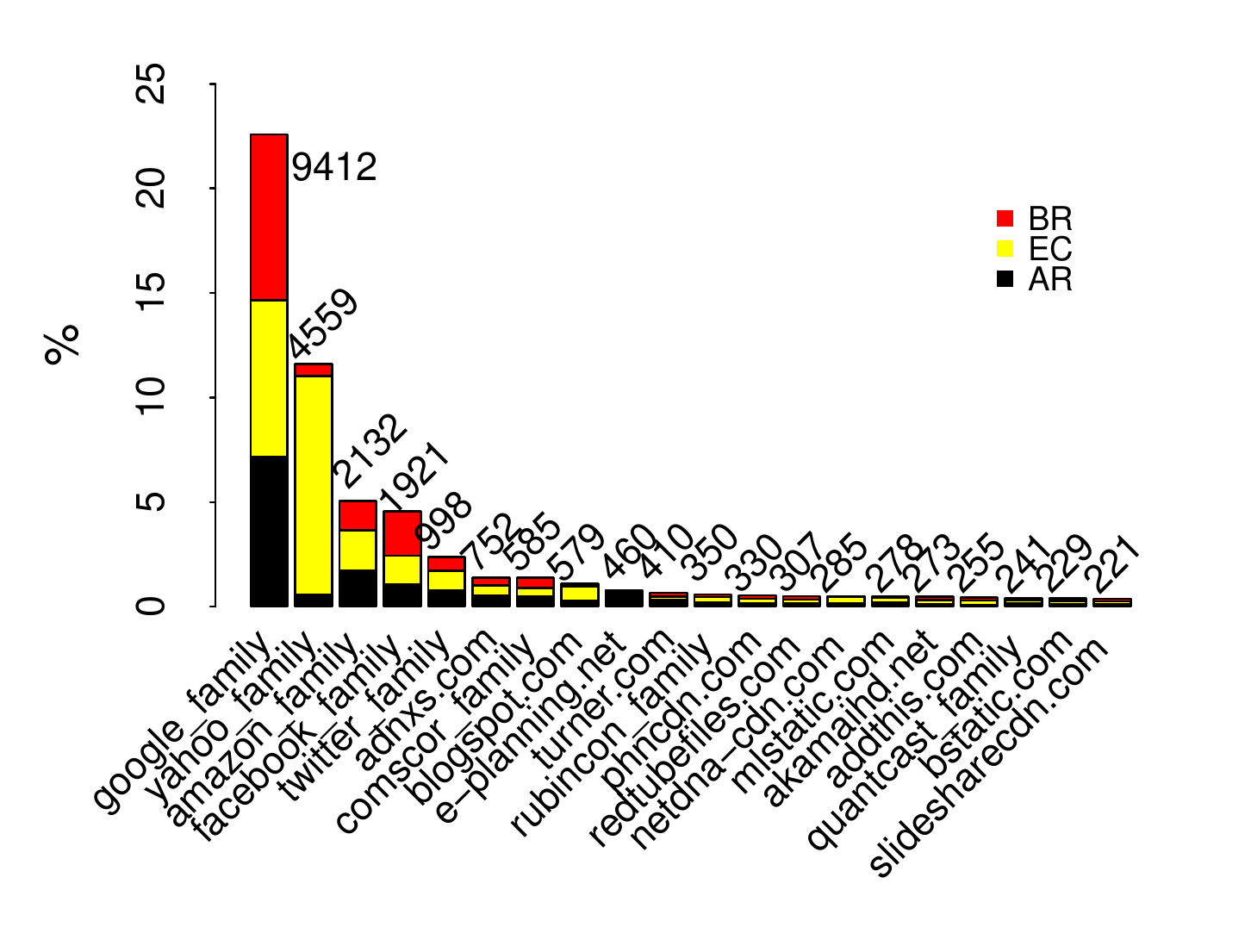}
                    \label{fig:toplatin}
  }

  \subfigure[Europe]{%
    \includegraphics[width=0.49\textwidth]
                    {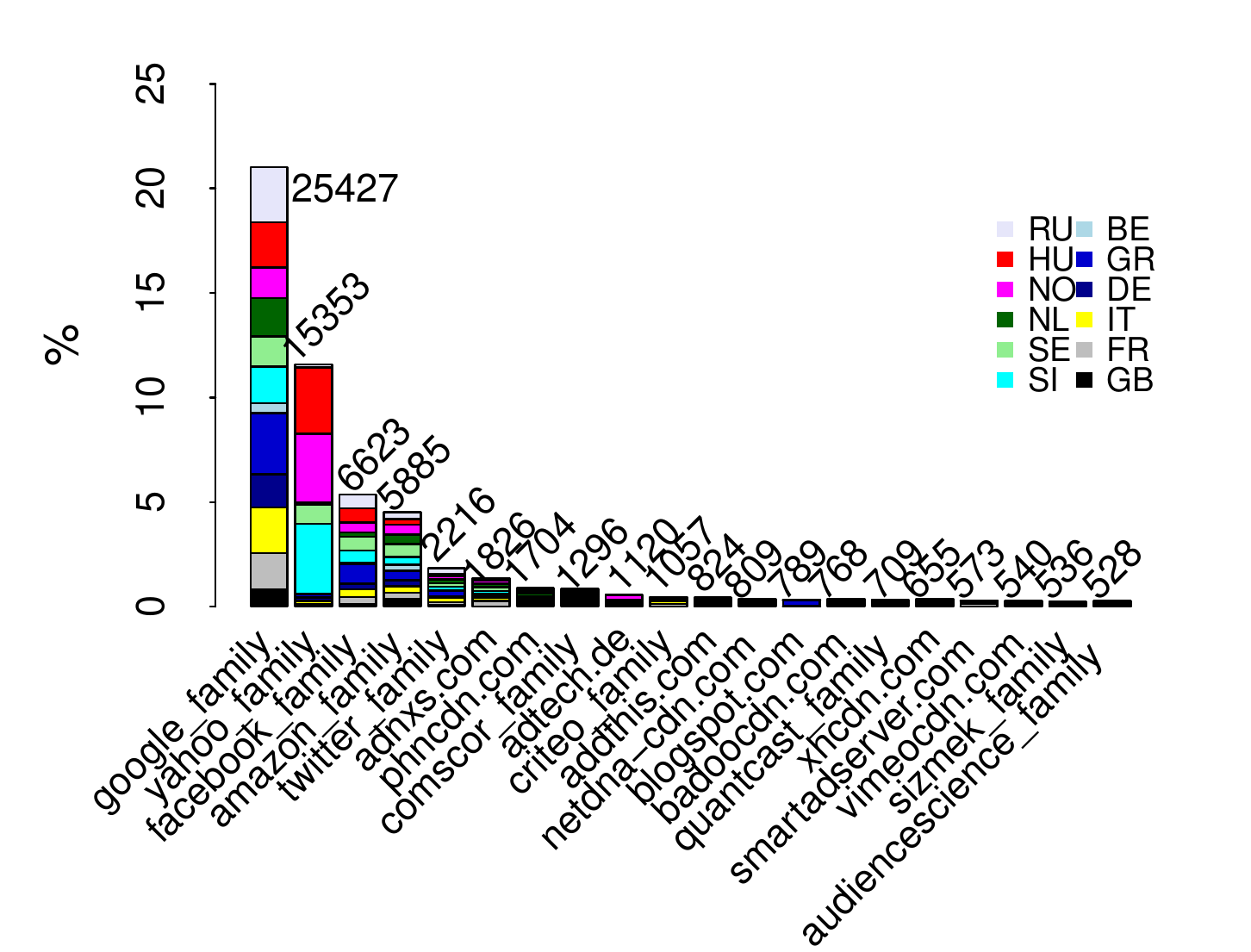}
                    \label{fig:topeu}
  }
  \subfigure[Oceania]{%
    \includegraphics[width=0.49\textwidth]
                    {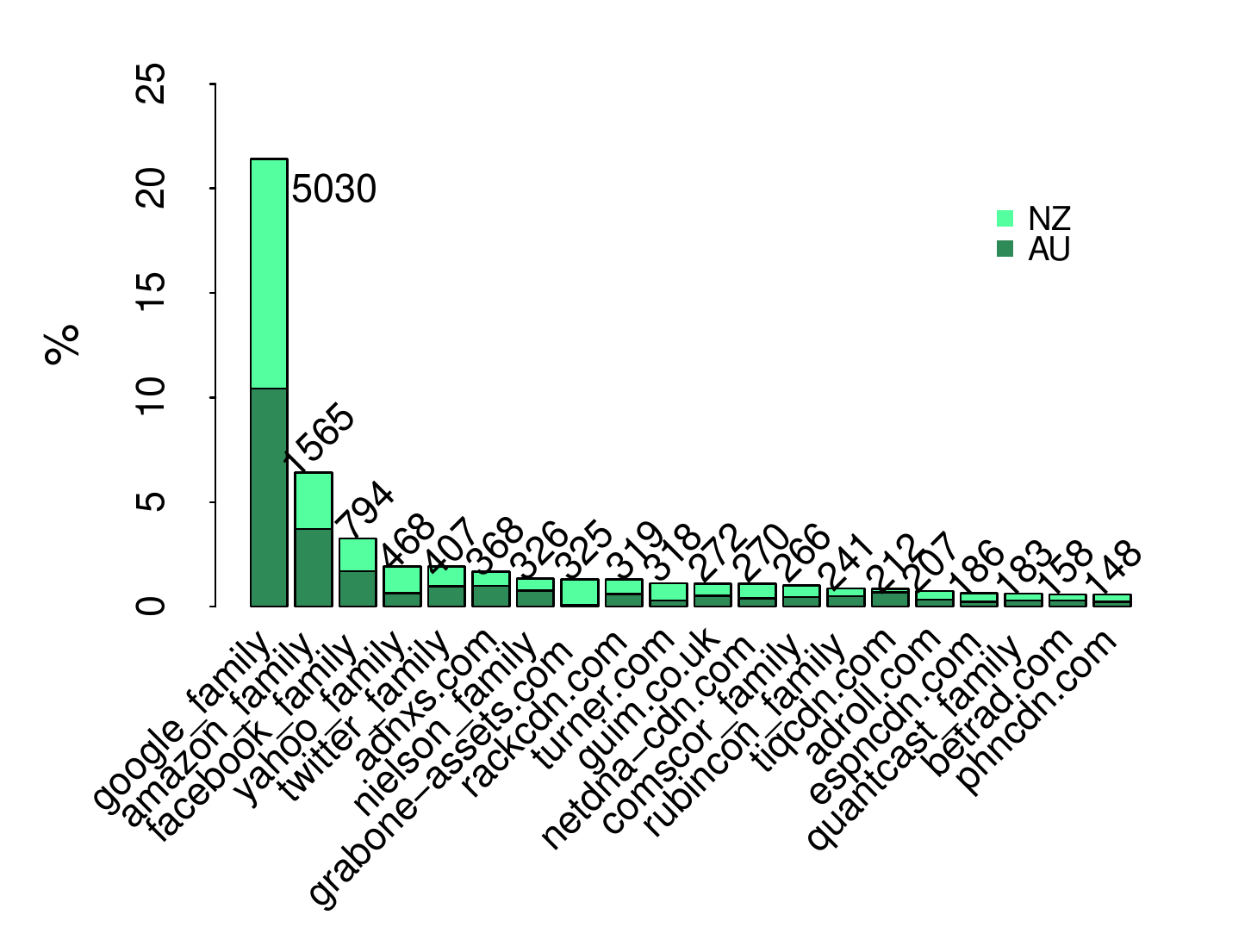}
                    \label{fig:topoc}
  }

  \subfigure[Middle-East]{%
    \includegraphics[width=0.49\textwidth]
                    {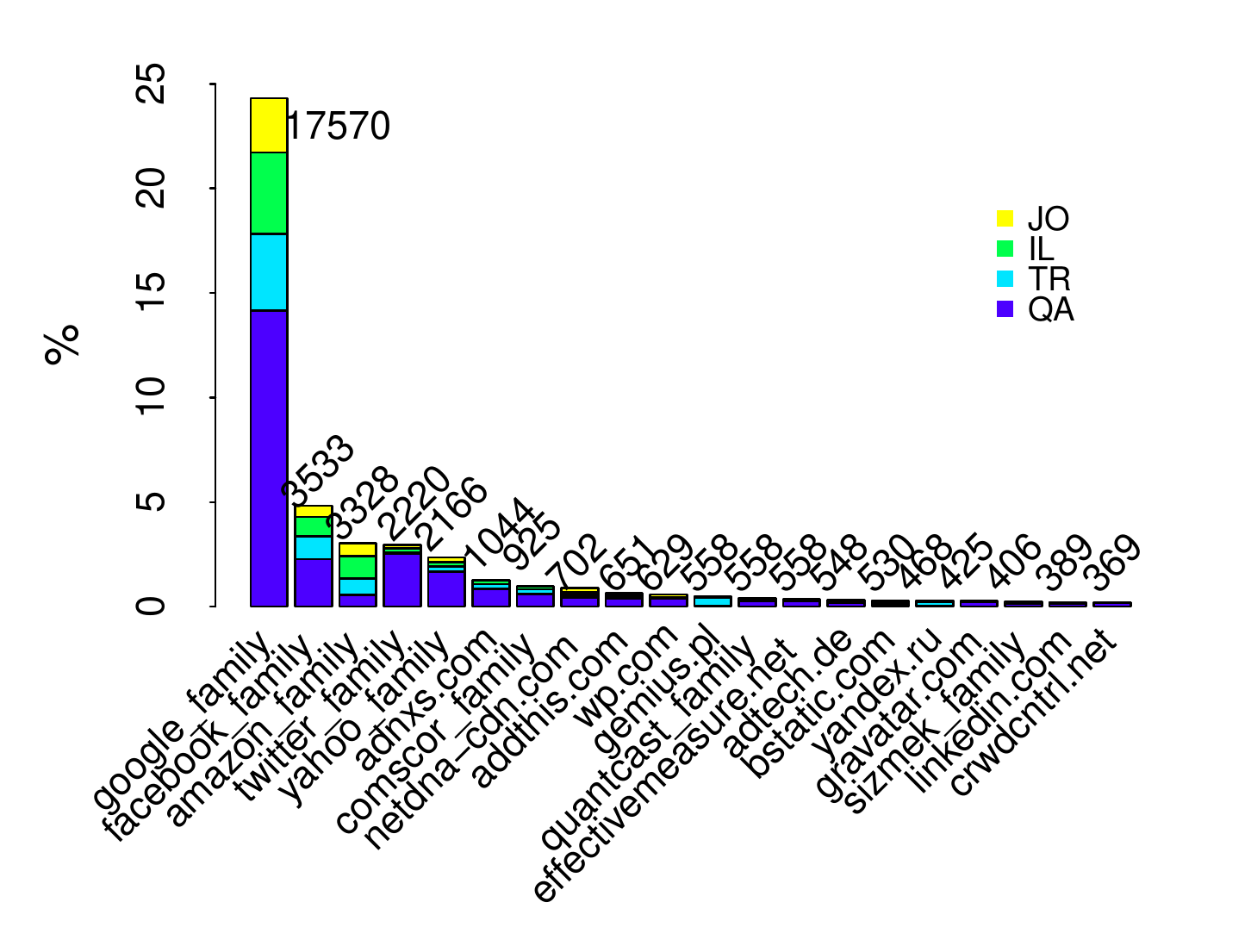}
                    \label{fig:topmd}
  }
  \hfill
  \subfigure[East-Asia]{%
    \includegraphics[width=0.49\textwidth]
                    {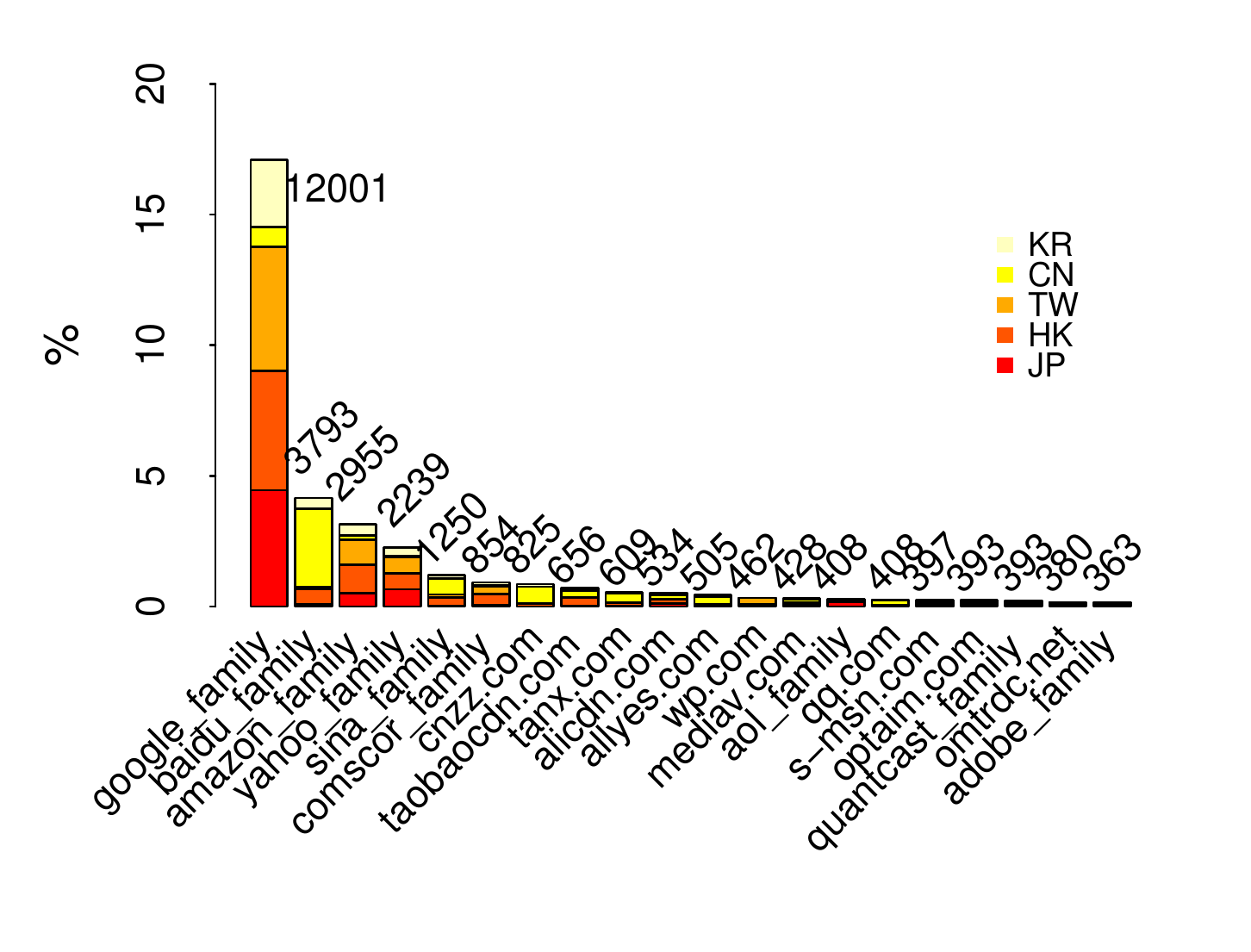}
                    \label{fig:topea}
  }

  \caption{\label{fig:top-20}Top-20 third-party websites by region. Occurrence count for each third-party is displayed above each bar.}
\end{figure*}

As well as the well-known services such as Google, we also observed some less well-known third-party services, spread across almost all regions. We present the top-20 in each region in Figure~\ref{fig:top-20}. We found third-party services belonging to Google, Amazon and Facebook roughly in the same position throughout our investigated regions (top four) while Yahoo, compared with other regions, has a notably higher position (second place) in Europe (Figure~\ref{fig:topeu}) and South America (Figure~\ref{fig:toplatin}). This difference in South America is due to the high number of occurrences of the Yahoo third-party requests in Ecuador (4304; 10\% of the third-party websites in South America). Similarly, Slovenia, Norway and Hungary contribute most in Europe (Figure~\ref{fig:topeu}).

Beside the most famous players, we identified other third-party services with extensive presence across all regions. For example, \url{scorecardresearch} belongs to comScore Inc., an analytics company, \url{netdna-cdn}, belongs to NetDNA, a CDN company, and \url{quantserve} belongs to QuantCast, a behavioural advertising company, almost appeared in all regions, except for the absence of \url{netdna-cdn} in East Asia. This presence implies a growing competitiveness of such businesses across the regions.

In addition to the global third-party services, we observed a notable presence of local third-parties in specific regions such as East Asia and Europe. We remind that local third-parties are those services which are physically located in the related region. In East Asia (Figure~\ref{fig:topea}), 11 cases from the top-20 are based in this region (e.g.,~\url{sina\-family}, \url{tabaocdn.com}); in Europe (Figure~\ref{fig:topeu}), 4 services amongst those presented are mainly found in European countries (DE-based: ~\url{adtech.de}; FR-based: \url{criteo.com},\url{smartadserver.com}; GB-based \url{badoocdn.com}). On the other hand, in Oceania and South America, there are far fewer local third-parties (one out of top-20) and in the Middle East there are none in the top-20.



\begin{table}
  \small
  \centering
  \renewcommand{\arraystretch}{1.2}
  \begin{tabular}{ l l }
    \textbf{TLD} & \textbf{Number}\\ \hline
    \url{.com}  & 3605 \\\hline
    Country Code & 1654 \\ \hline
    \url{.net}  & 793 \\\hline
    Other (e.g.,~\url{.biz}, \url{.asia}) & 444 \\ \hline
    \url{.org}  & 114 \\\hline
  \end{tabular}
  \caption{\label{tab:tld}The proportion of different Top Level Domain names of the third-parties.}
\end{table}

We next examine the ecosystem of third-party trackers in different regions to find out how this ecosystem looks like if we put the dominant and popular players aside. We also investigate ``unconventional" third-party services and their distribution in different regions. In our analysis, we excluded all those third-party websites that had a country code or a popular TLD name (com, org or net). This left us with about 4\%(= 235) of the total identified third-party trackers. Table~\ref{tab:tld} shows the proportion of different TLDs amongst the third-parties.



\begin{figure*}
  \centering

  \subfigure[North-America]{%
    \includegraphics[width=0.49\textwidth]
                    {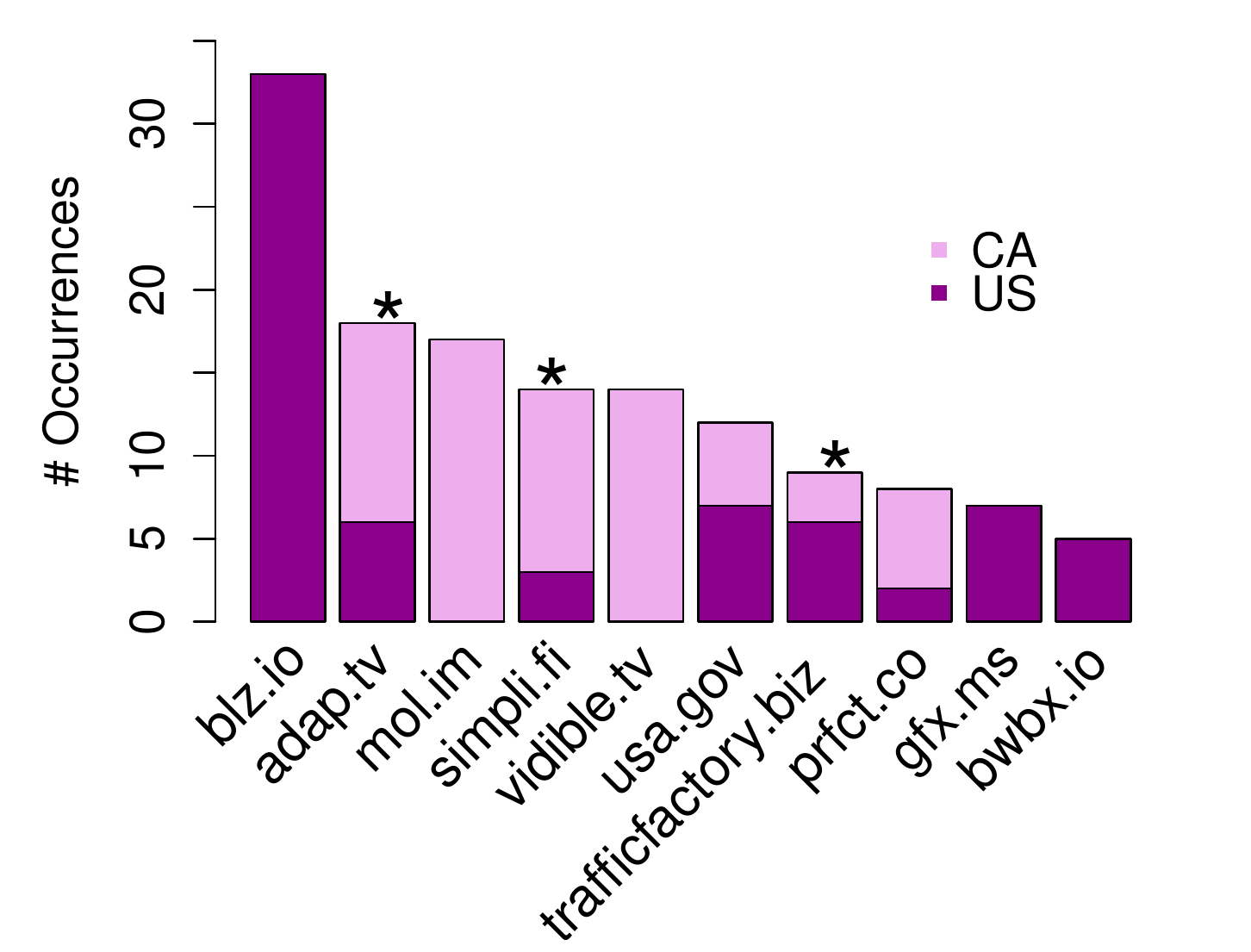}
                    \label{fig:us_nocountry}
  }
  \hfill
  \subfigure[South-America]{%
    \includegraphics[width=0.49\textwidth]
                    {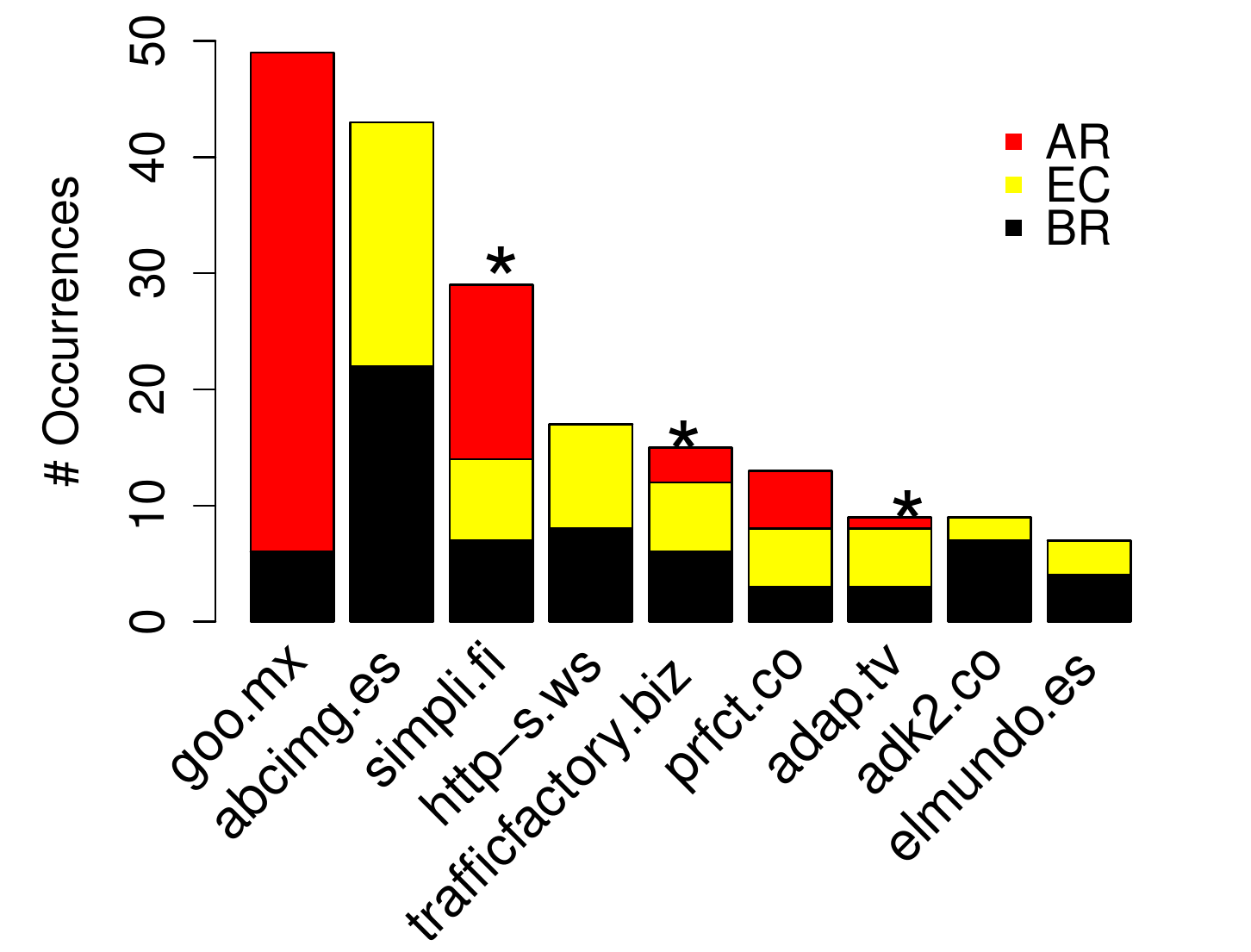}
                    \label{fig:latin_nocountry}
  }

  \subfigure[Europe]{%
    \includegraphics[width=0.49\textwidth]
                    {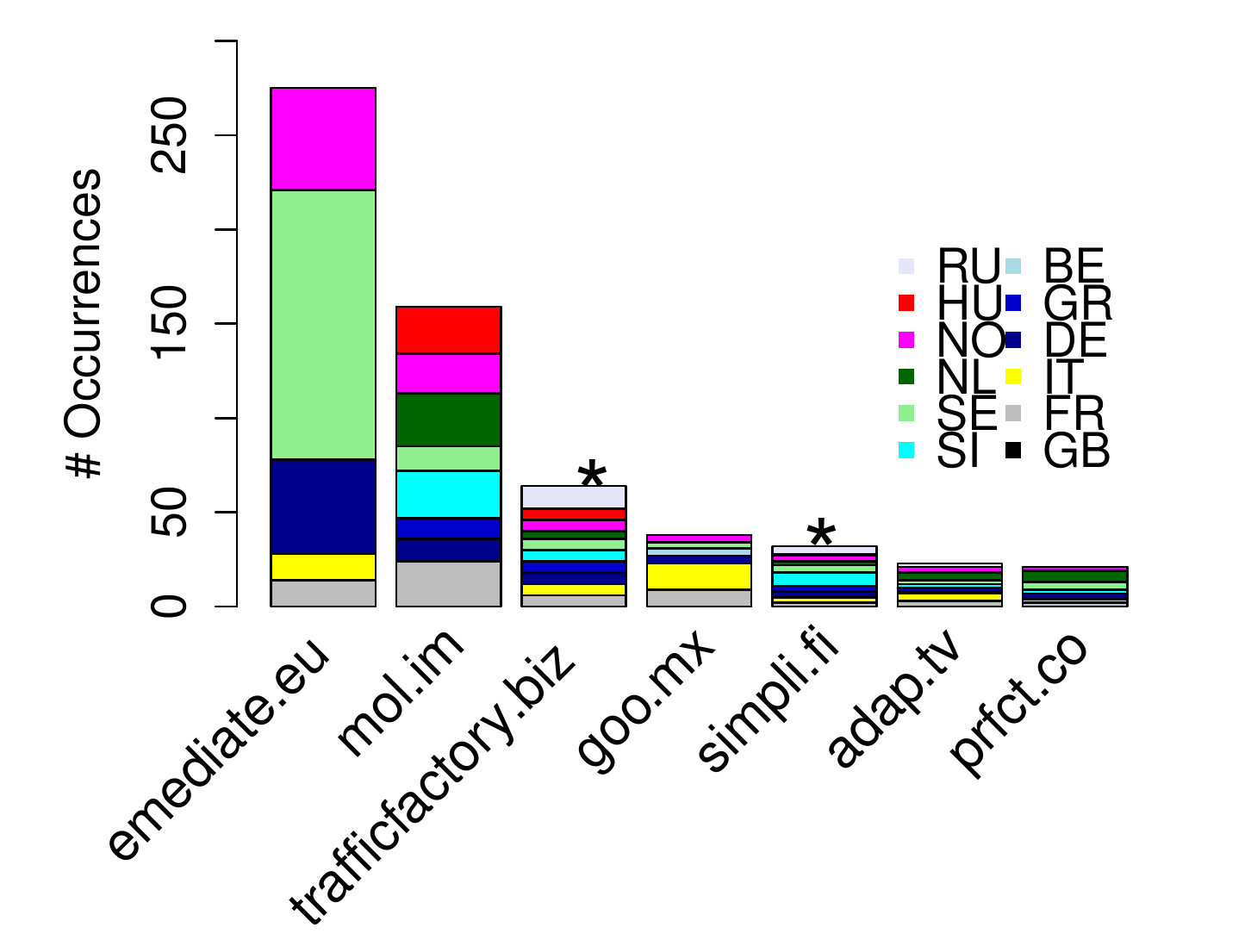}
                    \label{fig:eu_nocountry}%
  }
  \hfill
  \subfigure[Oceania]{%
    \includegraphics[width=0.49\textwidth]
                    {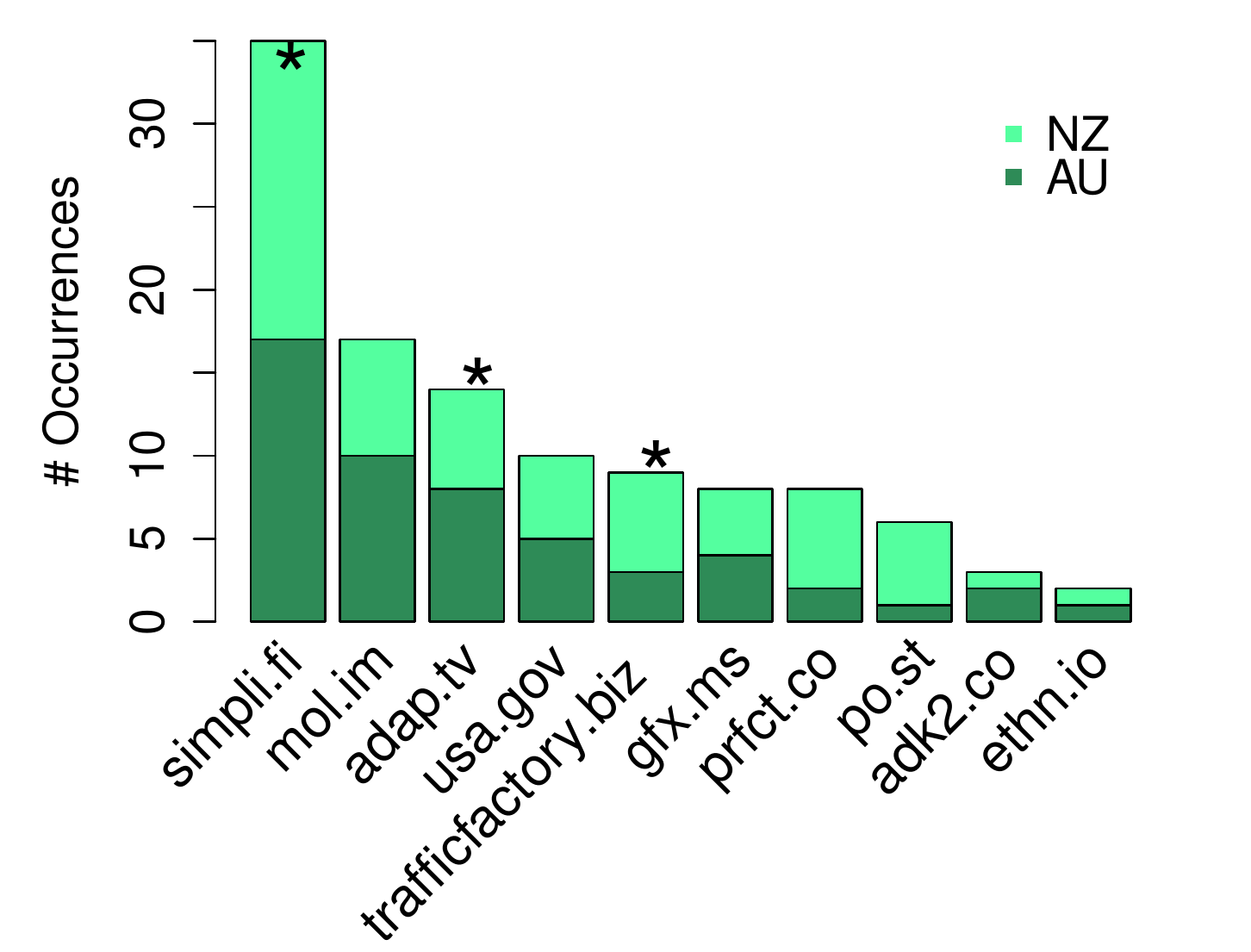}
                    \label{fig:oc_nocountry}
  }

  \subfigure[Middle East]{%
    \includegraphics[width=0.49\textwidth]
                    {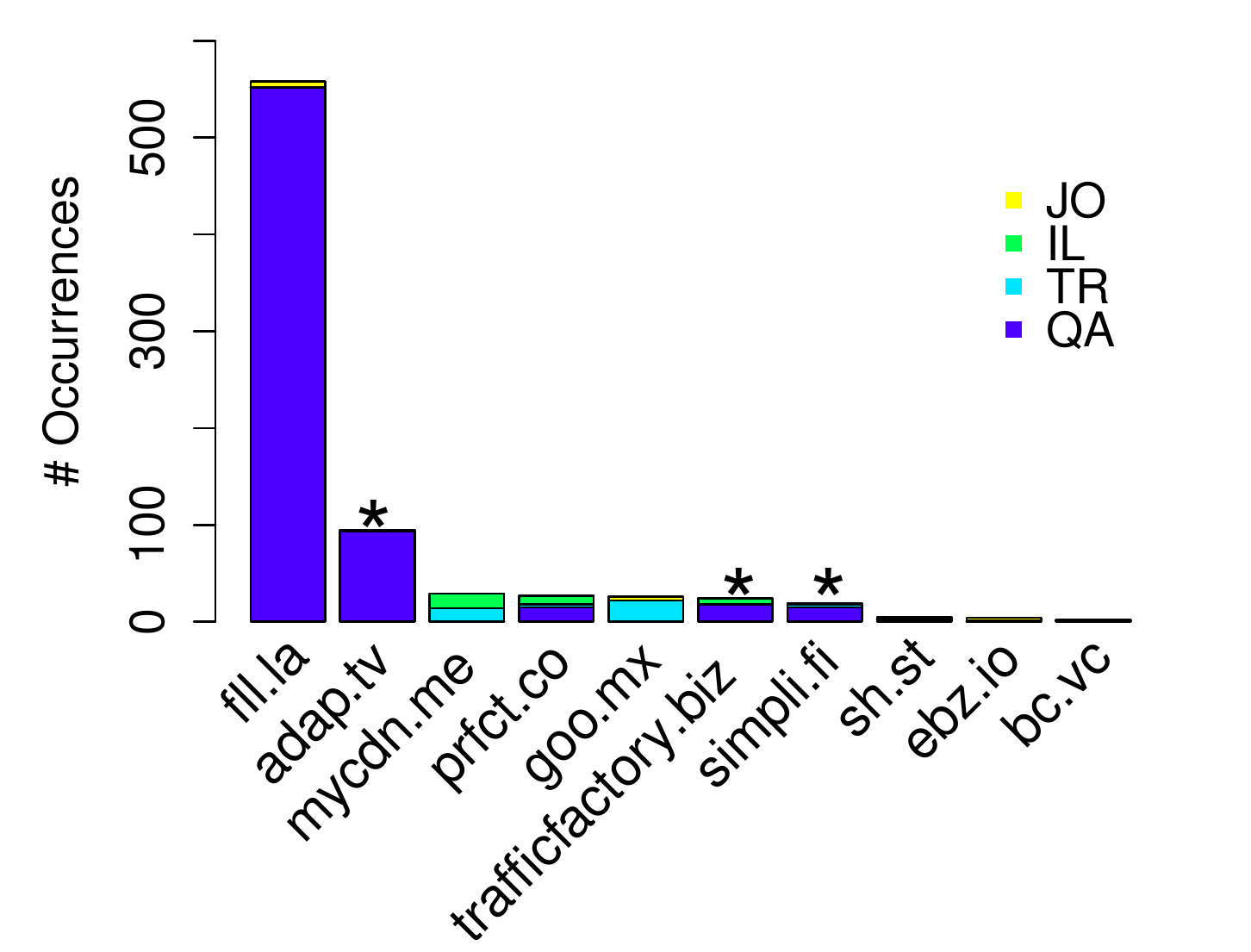}
                    \label{fig:md_nocountry}
  }
  \hfill
  \subfigure[East Asia]{%
    \includegraphics[width=0.49\textwidth]
                    {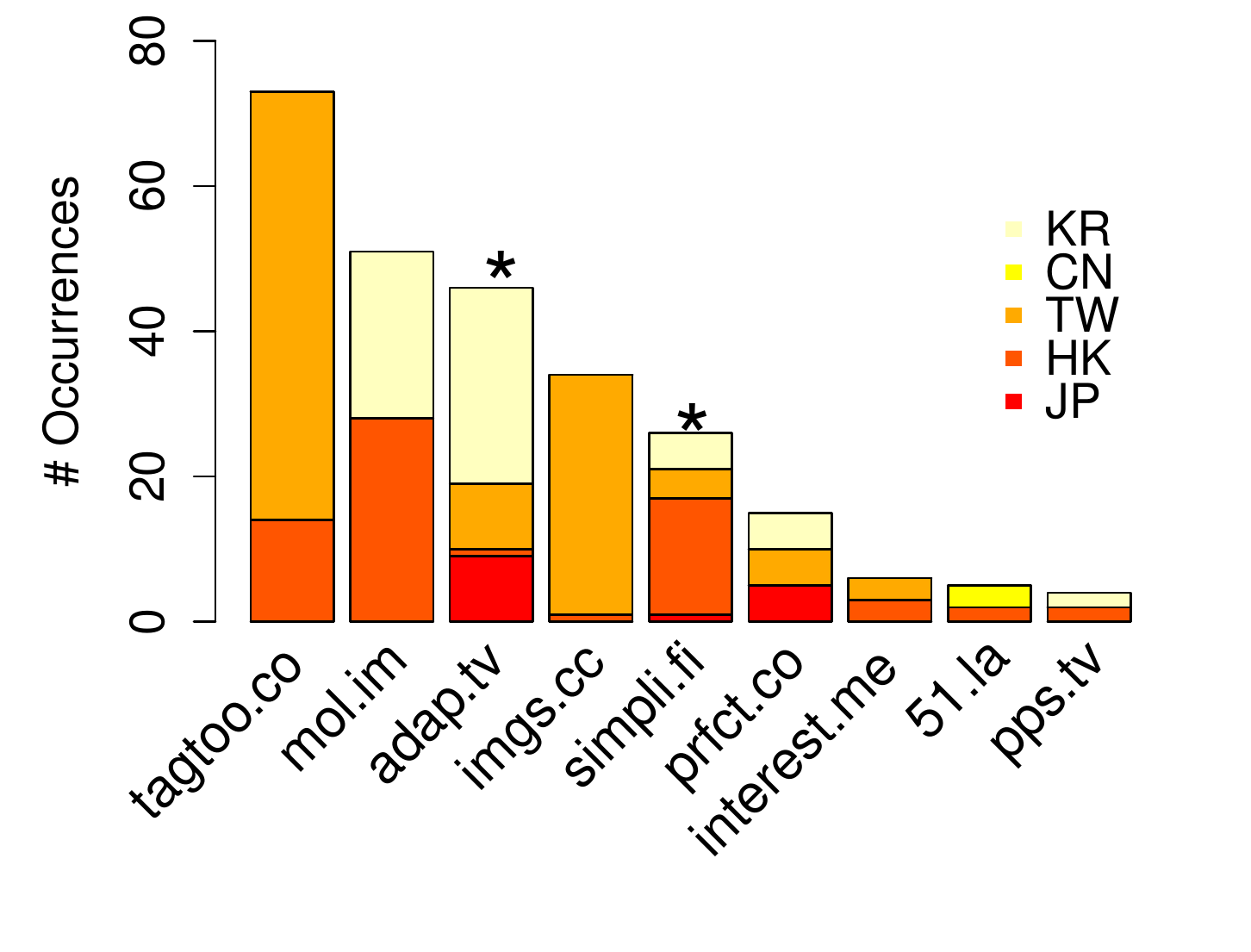}
                    \label{fig:es_nocountry}
  }

  \caption{\label{fig:nocountry}Top-10 small third-party websites in East Asia and Middle East. Globally observed sites are indicated by *.}
\end{figure*}

Figure~\ref{fig:nocountry} shows the top-10 atypical third-parties in each region, which appeared in over 50\% of the countries of that group. It is notable that amongst these atypical services, some are globally active. We identified three cases amongst the top ten: \url{simpli.fi} (a US-based ad tracker), \url{trafficfactory.biz} (a Netherlands-based advertising agency) and \url{adap.tv} (a US-based ad broker) that appeared in most of the regions of our dataset, denoted by (*) in the above figures.

Atypical services are almost equally spread across countries in Oceania (Figure~\ref{fig:oc_nocountry}), Europe (Figure~\ref{fig:eu_nocountry}) and South America (Figure~\ref{fig:latin_nocountry}), whereas in other regions the occurrence of small services is unequal amongst countries of each region. For example, in East Asia (Figure~\ref{fig:es_nocountry}) and the Middle East (Figure~\ref{fig:md_nocountry}), Taiwan and Qatar have high occurrence of the services in comparison with other countries in their own group.

In terms of region specific services, we identified a Denmark based ad serving third-party,\url{emediate.eu}, which mostly provide services in northern and central Europe such as Sweden, Norway, and Germany (Figure~\ref{fig:eu_nocountry}). We did not observe regional small services in other groups. However, the presence of some overseas services such as a Spain based web hosting, \url{abcimg.es}, in South America (Figure~\ref{fig:latin_nocountry}) and US governmental services such as \url{usa.gov} in Oceania (Figure~\ref{fig:oc_nocountry}) is of interest.
\section{Per-Country Analysis}
\label{sec:countries}

\begin{figure*}
  \subfigure[$x$-axis: country code \& $y$-axis: Alexa ranking]{%
    \includegraphics[width=0.32\textwidth]{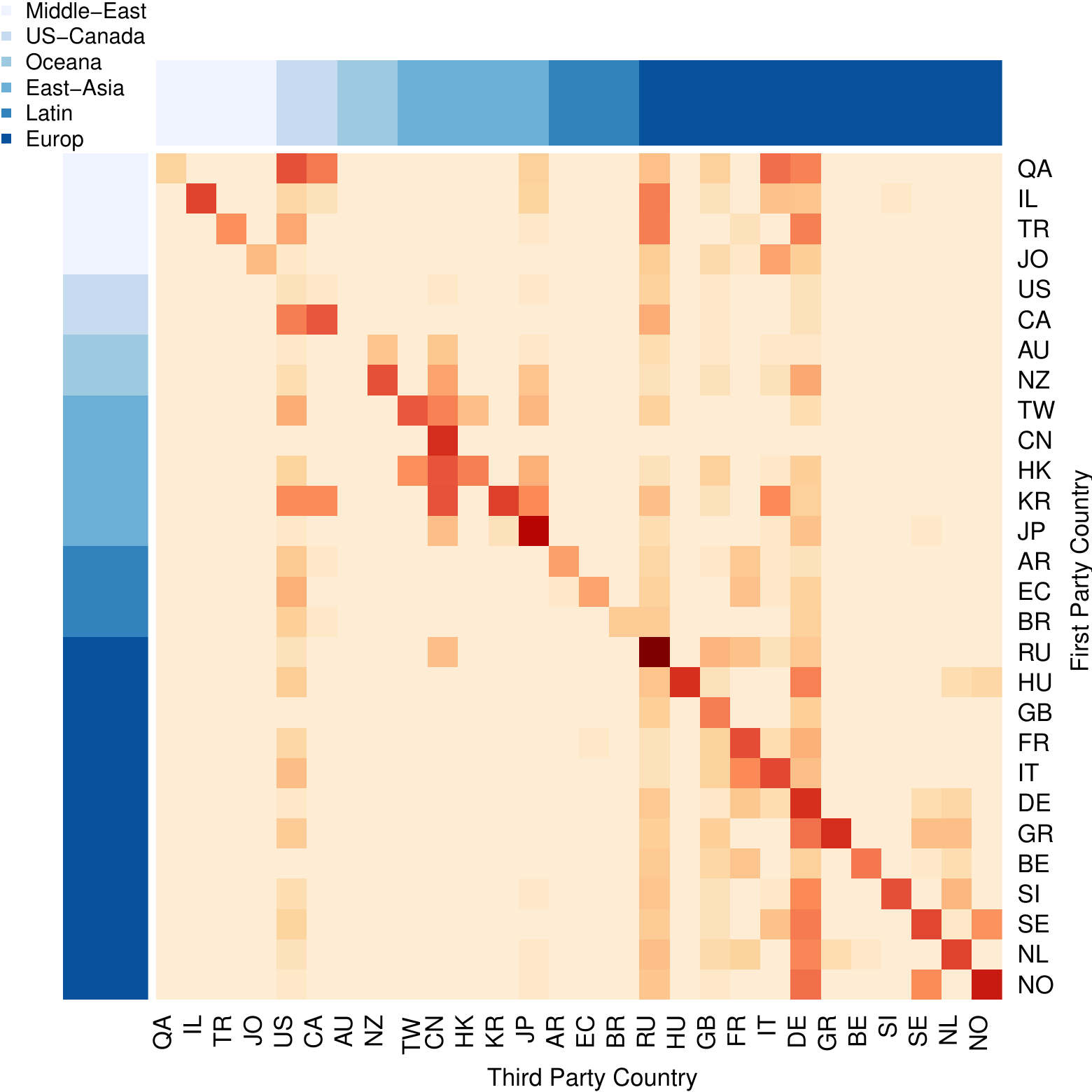}
    \label{map:cnt}
  }
  \hfill
   \subfigure[$x$-axis: physical location \& $y$-axis: Alexa ranking]{%
    \includegraphics[width=0.32\textwidth]{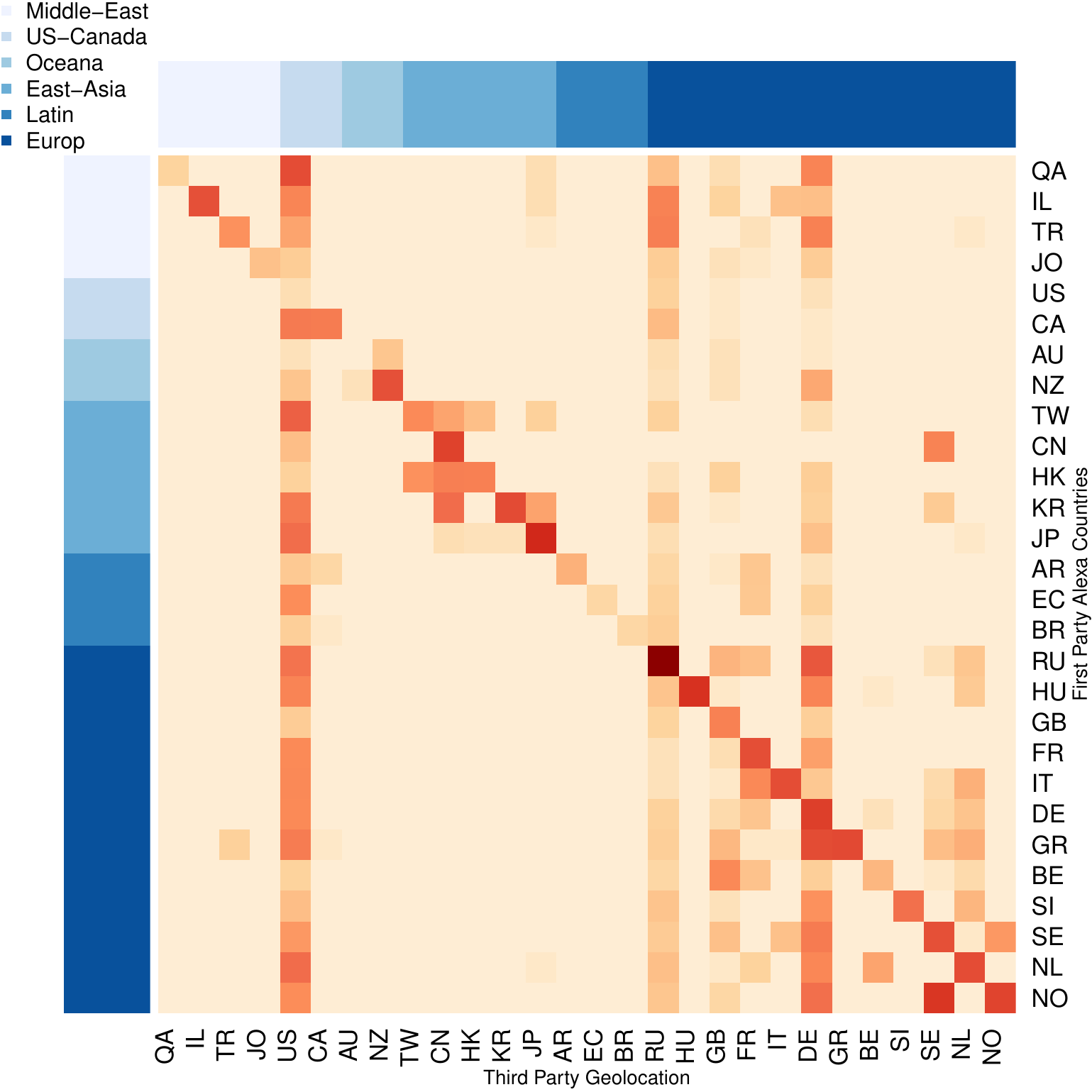}
    \label{map:geo_alexa}
  }
    \subfigure[$x$-axis \& $y$-axis: physical location]{%
    \includegraphics[width=0.32\textwidth]{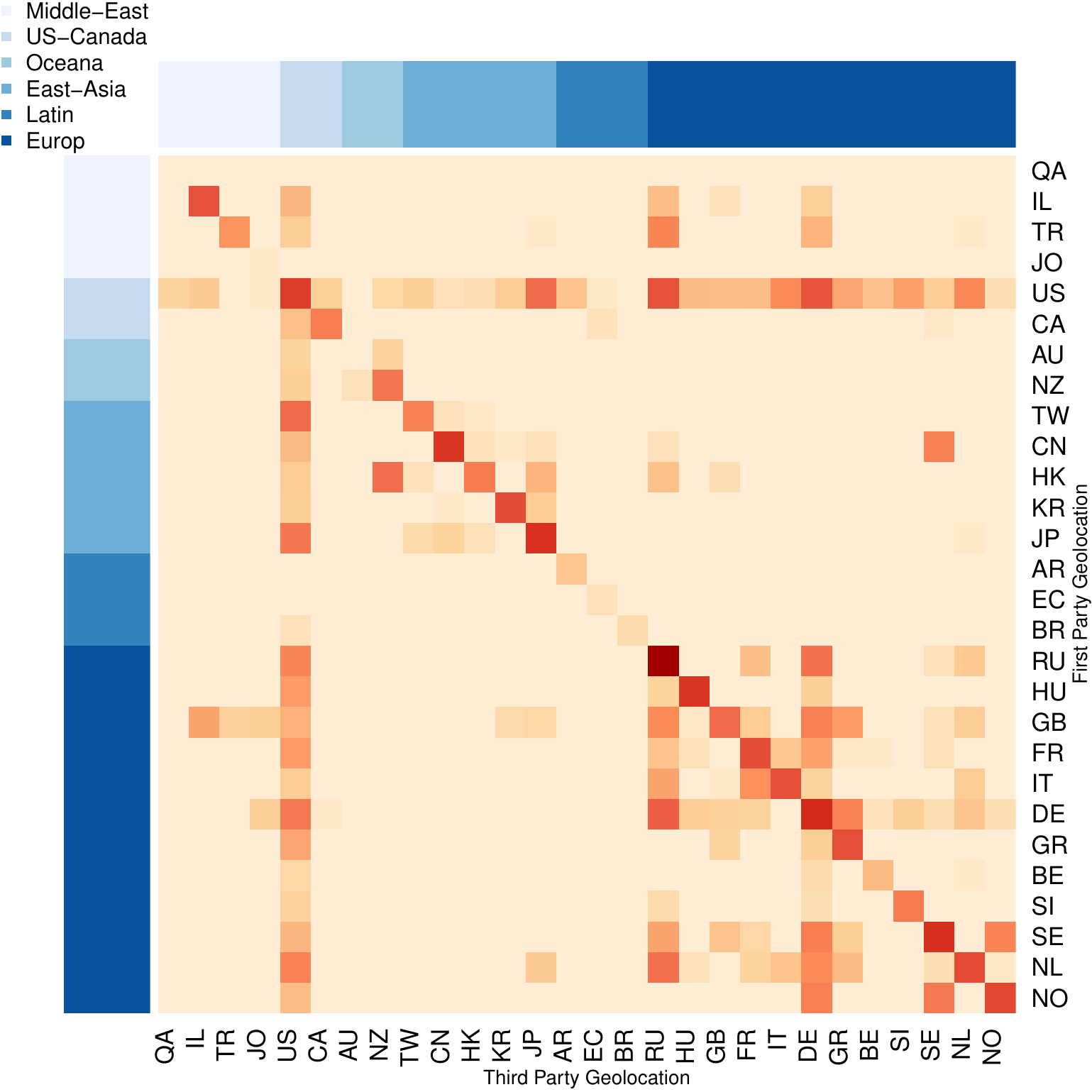}
    \label{map:geo_geo}
  }
  \caption{\label{map:heat_cnt}Heatmaps showing locations of third-parties. $y$-axis is the location of the first-party and $x$-axis is the location of the third-party. Darker colours indicate greater presence, and the region of each country in the two left-most plots is depicted by the colour of the blue bars on the left and at the top.}
\end{figure*}

In this section we look at the presence of the overseas third-party trackers in each country and region. We focus on a specific subset of third-parties which their TLD is a country code e.g., the TLD of ~\url{yadro.ru} is the country code of Russia, we refer to them as   \emph{locally named} third-parties. For this purpose, we exclude all third-parties with non Country Code TLD (CCTLD) as well as those which their CCTLD is not corresponding to the countries in our list. Our method is lower-bound for US-named third-parties because some TLDS such as \url{gov} is very popular for American domain names which we exclude them. We identified 1654 (24\%) unique third-parties with CCTLD amongst the other TLDs which are listed in the table~\ref{tab:tld}. 


We examine the penetration of locally named third-parties across popular websites of countries (based on top 500 Alexa ranking) where they do not belong to. The heat map in Figure~\ref{map:cnt} demonstrates such presence. The $y$-axis shows the country which visited website is popular in, while the $x$-axis corresponds to the CCTLD of third-parties which appeared in the visited sites. For example, the third row shows that the websites which are popular in Turkey embed some third-party trackers from Germany, Russia, United States and Turkey. In general, we found United States, Russia and Germany to be the countries with the locally named third-parties across popular websites of almost all countries in our dataset. Amongst other European countries, Norwegian and Sweden third-parties have a similar and notable presence in each other popular websites. One explanation for such presence is that a website can be popular in more than one country therefor its locally named third-party is considered as an overseas in other countries in which the website is popular. Another reason can stem from the difference between the country which a third-party CCTLD implies and the country where it is physically hosted. To investigate this possibility, we examined the correspondence between CCTLDs and the location of the third-parties. We found most of the CCTLDs corresponding to their actual physical location.  Heat map~\ref{map:geo_alexa} presents this result which is very similar to the heat map \ref{map:heat_cnt} except for those third-parties that their CCTDLs were not corresponding to their physical location. For instance locally named third-parties of Italy and Canada amongst popular websites of Qatar in the first row of figure\ref{map:cnt} are not present at the corresponding row in the figure \ref{map:geo_alexa} since they were not physically hosted in Italy and Canada. On the other hand, US-based locally named third-parties appear stronger which is due to the presence of some locally named third-parties which their TLD refers to another country rather than US. The presence of Sweden-based third-parties amongst popular websites in China is another revealed interesting point in this examination.

So far we observed the considerable presence of locally named third-parties  of some specific countries across popular websites of various other countries. We carry on our investigation to identify overseas third-parties using different approach, based on purely physical location of visited websites and locally named third-parties. So that we assign countries to the visited websites using their physically hosted location instead of where they are popular in as we did so far. In contrast to US, in countries of South America there is no overseas third-parties across the websites located there neither third-parties based in these countries being present across websites of other countries with slight exception for the websites based in North America. Similar to the previous examination, we identify the country of the locally named third-parties based on their physical location. The heat map of figure \ref{map:geo_geo} presents this result. Clearly, US has a unique situation; While across the majority of countries, with few exceptions, there are considerable number of locally named third-parties based in US, there are third-parties located in majority of countries which are embedded inside US-based websites. Similar to the both previous examinations, the presence of locally named third-parties hosted in Germany, North America and Russia is high amongst websites located in Israel and Turkey. However, we don\'t observe any third-party from these countries amongst websites located in the other countries of Middle East (Qatar and Jordan) neither amongst other regions except European countries and US in North American region. On the other hand we observe slight presence of third-parties hosted in Middle East countries except Qatar across websites located in Great Britain.  


We would like to clarify that the importance of presence of overseas third-parties is due to their possible access to the user\'s information from various countries who are visiting those popular websites. Considering various studies reporting the access of third-party trackers to the user's personally identifiable information such as full name, email or even very sensitive information including user's health condition, in addition to the growing trend of Web data surveillance by different governments, learning about countries which potentially have access to user information is helpful to understand the flow of data collection.    

\section{Discussion}
\label{sec:discussion}
In the previous section, we observed the presence of the overseas third-parties located in US, Germany and Russia amongst websites of almost all countries. This implies, not surprisingly, access to and potentially storage and processing of users' data from countries other than those where the first party service is hosted. We now briefly examine one hypothesis for this behaviour, that it is driven by the differing data protection and online privacy rules in key countries in those regions. Afterwards, we discuss the role of such regulation in the presence of third-party services.

The EU regulatory framework has a clear and comprehensive set of rules for data protection. This means that all businesses located in any members of the European Union should comply with such regulations. In terms of online privacy, according to the Directive 2002/58/EC and its amendment 2009/136/EC, known as ePrivacy Directive, websites which are using cookies or other technologies to collect user data should clearly inform users about such process and ask for their opt-in consent as soon as website is loaded on user's machine. Similarly, in Australia, any entity that collects personal information should notify user about that, however, the notification can be provided after actual data collection~\cite{au_law}. In US, in contrast to EU and Australia, there is no single, specific law that regulates the collection and use of personal data. Therefore, the related laws varies in different states as well as in different business sectors. The situation in countries such as China and Turkey becomes more ambiguous since there is no specific data protection law. We summarise the current regulation frameworks regarding data protection and international data transferring in the Table~\ref{tab:rules}.

Despite the existence of data protection regulations in many countries, and growing attention to the online privacy issues, the implementation and enforcement of such rules are not consistent with the laws in theory. For example, Germany has not implemented ePrivacy Directive, therefore, there is no requirement for active opt-in consent, e.g.,~by clicking on a pop-up window. It's suggested that browser cookie settings would remain adequate. While in UK having user opt-in consent is mandatory. In Russia, data protection rules have many similarities with the EU directives, however, the inconsistencies and complexity in the regulations lead to very limited enforcement of law. In US, on the other hand, with totally different approach of jurisdiction system (relying more on self-regulatory and guidelines) monitoring businesses is more complicated. These differences suggest the effect of implementation and enforcement of law on highly presence of third-party services in specific countries like US, Germany and Russia.

We should note that other factors such as technological advancement and political approaches could be influential in this field. Germany act as a commercial hub in Europe as well as owning the most advanced digital infrastructure such as DE.CEX, the largest Internet exchange point in the world. Russia based third-parties in Asia and Middle East is aligned with the general doctrine of the Russian government to broaden relationship with these regions~\cite{russia_news}. 

{
  \newcolumntype{g}{>{\columncolor{Gray}}p{0.110\textwidth}}
  \newcolumntype{w}{>{}p{0.110\textwidth}}
  \newcolumntype{h}{>{\columncolor{Gray}}c}
  \newcolumntype{v}{>{}c}
  \begin{table*}[t]
    \small
    \renewcommand{\arraystretch}{1.2}
    \setlength{\tabcolsep}{3pt}
    \centering

    \begin{tabular}{ g w g w g w g w }
      \multicolumn{8}{v}{\ding{52} Yes \ding{54} No \ding{119} Partial} \\
      \hline
      \multicolumn{1}{h}{\bf Question}
      & \multicolumn{1}{v}{\bf US}
      & \multicolumn{1}{h}{\bf DE}
      & \multicolumn{1}{v}{\bf UK}
      & \multicolumn{1}{h}{\bf AU}
      & \multicolumn{1}{v}{\bf RU}
      & \multicolumn{1}{h}{\bf CN}
      & \multicolumn{1}{h}{\bf TR} \\
      \hline\hline
      Existence of data protection laws & \ding{119} & \ding{52} & \ding{52} & \ding{52} & \ding{52} & \ding{119} & \ding{54} \\\hline
      Coverage of privacy law & Sectoral & Comprehensive & Comprehensive & Comprehensive & Comprehensive & Sectoral & Not applicable \\\hline
      Effective regulator to enforce the privacy laws & Sectoral regulation & Sectoral regulation & National regulation & National regulation & National regulation & None & None \\\hline
      Cookie specific regulation & \ding{54} & \ding{52} & \ding{52} & \ding{54} & \ding{54} & \ding{54} & \ding{54} \\\hline
      Dealing with non-essential cookies & Informing user via site policy (guideline) & Opt-out mechanism (regulation) & Opt-in (regulation) & Notifying user before or after visiting a site & None & None & None \\\hline

      Overseas transferal of data & US entities are liable for adequate protect of data through security safeguards, protocols or contractual model, with different industrial sectors having different regulations & The data recipient must ensure an adequate level of data protection & Adequacy assessment of data protection law in the relevant country (outside EEA) is required or the organisation must get their Binding Corporate Rules approved by the national Information Commissioner & Entity must take ``reasonable steps'' to ensure the principles are not breached overseas, e.g.,~if a cloud service provider is planning on sending data overseas, it should have a contract in place to make sure data will not be misused & Data can be transferred to Strasbourg Convention states or other states that ensure adequate protection of personal data & No specific regulation; based on the nature of data there are certain industrial regulations, e.g.,~information collected by commercial banks is not allowed to be transferred overseas & No specific regulation other than requires consent, though individual cases may have additional requirements \\\hline
    \end{tabular}
    \caption{\label{tab:rules}Comparison of data protection and data transferring across different countries.}
  \end{table*}
}
\section{Related work}
\label{sec:related}

A number of studies have analyzed third-party trackers from different points of view. Krishnamurthy \& Wills~\cite{Krishnamurthy} investigated the expansion of third-party trackers from 2005 for a period of three years. They showed how tracking has changed with time and the acquisitions of various companies. They had previously analysed the growing association between first-party and third-parties in~\cite{Krishnamurthy:footprint}. In~\cite{krishnamurthy2011privacy}, they examined the access of third-parties to personal information based on the category of the first-party website in which they are embedded. They found that websites providing health and travel-related services disclose more information to third-parties than other types of websites.

Roesner et al.~\cite{roesner2012detecting} proposed a framework for classifying the behaviour of web trackers based on the scope of the browsing profile they produce. They show the spread of the identified classes amongst the top 500 websites in the world. Gomer et al~\cite{natasa:tracking_network} focused on the network aspects of third-party trackers in three search markets. 
They show a consistent network structure across different markets as well as high efficiency in exchanging information among third-parties. Mayer et al~\cite{Mayer:survey:2012} surveyed different techniques which are used by web trackers to collect user information.

While the above studies focused on the technical capabilities of specific types of third-party trackers, our study examines the presence of all third-parties across different regions of the world. Kulshrestha et al.~\cite{twitter} show the way in which users in the various parts of the world have different (local and global) interaction on the Twitter social network. Our work is closest to that of Castelluccia et al.~\cite{castellucia:hal-00832784}, which analyse the top 100 most popular sites worldwide across a number of countries, to assess their tracker behaviours. They focused on measuring the penetration of US-based services in different countries, whereas, we, focus on the regional presence of third-party trackers as well as less-known cases.

From a privacy point of view, Ur et al~\cite{Ur:perception:2012} report users' strong concerns about data collection done by ad trackers. Moreover, Bellman et al.~\cite{Bellman:privacy_concerns:2004} showed the significant effect of culture and national regulation on users' privacy concerns and consequently suggests localized privacy policies. In our work, we also show the impact of regional characteristics on the structure of the third-party ecosystem, and suggest to further investigate how privacy policies in this ecosystem are affected by the regional regulatory frameworks in place.


\section{Conclusions}
\label{sec:conclusion}

In this paper, we had presented a study of the geographic differences in the third-party ecosystem. We sampled the Alexa top-500 most popular websites in each of 28 countries across widely spread regions of the world: North America, South America, Europe, East Asia, Middle East and Oceania. We examined the global and regional presence of the large, small and atypical third-party services in each region, as well as their penetration into other regions. We exposed connections amongst countries within each region by examining the geographic presence of third-parties that serve a big share of today's web.

Unsurprisingly, we found overall that a small number of international corporations are heavily dominant in all countries and regions. We observed significant differences in the numbers of observed third-parties across regions, with Turkey and Israel in the Middle East region standing out as having considerably more local third-parties than most countries. We observed considerably greater regional dominance of third-parties in Europe and East Asia, perhaps indicating greater commercial collaboration among companies in countries in those regions. In contrast, presence countries in North America is dominated by local third-parties.

We hope that the findings of our study will help better understand the international trade of personal information and accordingly adapt privacy protection solutions. Indeed, we highlighted the potential influence of regulatory constraints on the presence of third-parties. One example is Russia where the complexity and ambiguity of privacy regulation limits their implementation. Our observations suggest that privacy regulation, particularly in the area of cloud computing, requires more attention from the regulatory community.

As further work, we would like to expand the analysis by examining the relationships between the third-party and the peering ecosystems, especially to shed light on the physical presence and deployment of third-parties. Another interesting further work is to examine in details the type of services provided by third-parties in various regions of the world. Finally, we need to find some way to collect data about these relationships in Africa.

\section*{Acknowledgements}
We acknowledge constructive feedback and advice from Balachander Krishnamurthy. This work was funded in part by Horizon Digital Economy Research, RCUK grant EP/G065802/1.

{
  \bibliographystyle{unsrt}
  \bibliography{marjan}

\begin{thebibliography}{10}

\bibitem{vallina2012breaking}
Narseo Vallina-Rodriguez, Jay Shah, Alessandro Finamore, Yan Grunenberger,
  Konstantina Papagiannaki, Hamed Haddadi, and Jon Crowcroft.
\newblock Breaking for commercials: characterizing mobile advertising.
\newblock In {\em Proceedings of the ACM Internet Measurement Conference
  (IMC)}, 2012.

\bibitem{Krishnamurthy}
Balachander Krishnamurthy and Craig Wills.
\newblock Privacy diffusion on the web: a longitudinal perspective.
\newblock In {\em Proceedings of the 18th international conference on World
  Wide Web (WWW)}, pages 541--550, New York, NY, USA, 2009. ACM.

\bibitem{marjanTMA}
Marjan Falahrastegar, Hamed Haddadi, Steve Uhlig, and Richard Mortier.
\newblock The rise of panopticons: Examining region-specific third-party web
  tracking.
\newblock In Alberto Dainotti, Anirban Mahanti, and Steve Uhlig, editors, {\em
  Traffic Monitoring and Analysis}, volume 8406 of {\em Lecture Notes in
  Computer Science}, pages 104--114. Springer Berlin Heidelberg, 2014.

\bibitem{nsa}
{NSA using Google's online ad tracking tools to spy on web users}.
\newblock
  \url{http://www.computing.co.uk/ctg/news/2318698/nsa-using-googles-online-ad-/tracking-tools-to-spy-on-web-users}.

\bibitem{castellucia:hal-00832784}
Claude Castellucia, Stephane Grumbach, and Lukasz Olejnik.
\newblock {Data Harvesting 2.0: from the Visible to the Invisible Web}.
\newblock In {\em {The 12th Workshop on the Economics of Information
  Security}}, Washington, DC, USA, June 2013.

\bibitem{Krishnamurthy:summer:2010}
Balachander Krishnamurthy.
\newblock I know what you will do next summer.
\newblock {\em SIGCOMM Comput. Commun. Rev.}, 40(5):65--70, October 2010.

\bibitem{Krishnamurthy:footprint}
Balachander Krishnamurthy and Craig~E. Wills.
\newblock Generating a privacy footprint on the {Internet}.
\newblock In {\em Proceedings of the 6th ACM SIGCOMM conference on Internet
  measurement}, IMC '06, pages 65--70, New York, NY, USA, 2006. ACM.

\bibitem{planetlab}
PlanetLab.
\newblock Planetlab: An open platform for developing, deploying and accessing
  planetary-scale services.
\newblock \url{https://www.planet-lab.org/}.

\bibitem{webindex}
Web index.
\newblock \url{http://thewebindex.org/}.

\bibitem{collusion}
Collusion~Firefox add on.
\newblock Collusion firefox add-on.
\newblock \url{http://collusion.toolness.org/}.

\bibitem{au_law}
Australian privacy principle 5 — notification of the collection of personal
  information.
\newblock
  \url{http://www.oaic.gov.au/images/documents/privacy/engaging-with-you/current-privacy-consultations/Draft-APP-Guidelines-2013/Draft_APP_Guidelines_Chapter_5__APP_5.pdf
  }.

\bibitem{russia_news}
Richard Connolly.
\newblock The other pivot to asia and why success in china is not all it seems
  for putin’s russia.
\newblock
  \url{https://theconversation.com/the-other-pivot-to-asia-and-why/-success-in-china-is-not-all-it/-seems-for-putins-russia-27039/}.

\bibitem{krishnamurthy2011privacy}
Balachander Krishnamurthy, Konstantin Naryshkin, and Craig Wills.
\newblock Privacy leakage vs. protection measures: the growing disconnect.
\newblock In {\em Proceedigs of the Web 2.0 Security and Privacy Workshop},
  2011.

\bibitem{roesner2012detecting}
Franziska Roesner, Tadayoshi Kohno, and David Wetherall.
\newblock Detecting and defending against third-party tracking on the web.
\newblock In {\em USENIX Symposium on Networking Systems Design and
  Implementation (NSDI)}. USENIX, 2012.

\bibitem{natasa:tracking_network}
Richard Gomer, Eduarda~Mendes Rodrigues, Natasa Milic-Frayling, and M.C.
  Schraefel.
\newblock Network analysis of third party tracking: User exposure to tracking
  cookies through search.
\newblock {\em Web Intelligence and Intelligent Agent Technology, IEEE/WIC/ACM
  International Conference on}, 1:549--556, 2013.

\bibitem{Mayer:survey:2012}
Jonathan~R. Mayer and John~C. Mitchell.
\newblock Third-party web tracking: Policy and technology.
\newblock In {\em Proceedings of the 2012 IEEE Symposium on Security and
  Privacy}, SP '12, pages 413--427, Washington, DC, USA, 2012. IEEE Computer
  Society.

\bibitem{twitter}
Juhi Kulshrestha, Farshad Kooti, Ashkan Nikravesh, and Krishna~P. Gummadi.
\newblock {Geographic Dissection of the Twitter Network}.
\newblock In {\em In Proceedings of the 6th International AAAI Conference on
  Weblogs and Social Media (ICWSM)}, Dublin, Ireland, June 2012.

\bibitem{Ur:perception:2012}
Blase Ur, Pedro~Giovanni Leon, Lorrie~Faith Cranor, Richard Shay, and Yang
  Wang.
\newblock Smart, useful, scary, creepy: Perceptions of online behavioral
  advertising.
\newblock In {\em Proceedings of the Eighth Symposium on Usable Privacy and
  Security}, SOUPS '12, pages 4:1--4:15, New York, NY, USA, 2012. ACM.

\bibitem{Bellman:privacy_concerns:2004}
Steven Bellman, Senior Lecturer, Eric~J. Johnson, Stephen~J. Kobrin, William~H.
  Wurster, Professor~Multinational Management, and Gerald~L. Lohse.
\newblock G.: International differences in information privacy concerns: A
  global survey of consumers.
\newblock {\em The Information Society}, pages 313--324, 2004.

\end{thebibliography}
}

\appendix
\section*{Appendix A}
\label{App:appendix}
The full list of the identified family companies is shown in table~\ref{tab:aqn}.

{
  \newcolumntype{g}{>{\columncolor{Gray}}p{0.140\textwidth}}
  \newcolumntype{w}{>{}p{0.140\textwidth}}
  \newcolumntype{h}{>{\columncolor{Gray}}c}
  \newcolumntype{v}{>{}c}
  \begin{table*}[t]
    \renewcommand{\arraystretch}{1.2}
    \setlength{\tabcolsep}{3pt}
    \centering
    \small

    \begin{tabular}{ g w g w g g}
      \hline      
      \multicolumn{6}{h}{\bf Google}      
      \\
      \hline
      \multicolumn{1}{h}{\bf DoubleClick}
      & \multicolumn{1}{v}{\bf YouTube}
      & \multicolumn{1}{h}{\bf Blogger}
      & \multicolumn{1}{v}{\bf Other Acqn..}
      & \multicolumn{2}{h}{\bf Google Specific}
      \\
      \hline\hline

      doubleclick.net	&	youtube.com	&	blogblog.com	&	android.com	&	ajax.googleapis.com	&	{googlesyndication\par\hfill.com}	\\\hline
      doubleclick.com	&	{youtube.googleapis\par\hfill.com}	&	blogger.com	&	{channelintelligence\par\hfill.com}	&	{content.googleapis\par\hfill.com}	&	{googletagmanager\par\hfill.com}	\\\hline
      2mdn.net	&	{youtube-nocookie\par\hfill.com}	&	{blogger-comments\par\hfill.googlecode.com}	&	eedburner.com	&	{fonts.googleapis.com}	&	{googleusercontent\par\hfill.com}	\\\hline
      ytimg.com	&		&	blogspot.com	&	gmodules.com	&	goo.gl	&	googlevideo.com	\\\hline
      &		&	{wordtechnews\par\hfill.blogspot.com}	&	invitemedia.com	&	{googleadservices\par\hfill.com}	&	gstatic.cn	\\\hline
      &		&	ggpht.com	&	orkut.com	&	googleadsserving.cn	&	gstatic.com	\\\hline
      &		&		&	recaptcha.net	&	{google-analytics.com}	&	{javaplugins\par\hfill.googlecode.com}	\\\hline
      &		&		&	urchin.com	&	googleapis.com	&	{maps.googleapis.com}	\\\hline
      &		&		&		&	googlecode.com	&	{translate.googleapis\par\hfill.com}	\\\hline
      &		&		&		&	google.com	&	{www.googleapis.com}	\\\hline
      &		&		&		&		&	{googlecommerce.com}	\\\hline
      \hline
    \end{tabular}
    \begin{tabular}{ g w g w w w }
      \multicolumn{6}{h}{\bf AOL}
      \\
      \hline
      \multicolumn{1}{h}{\bf Advertising.com}
      & \multicolumn{1}{v}{\bf Huffington Post}
      & \multicolumn{1}{h}{\bf Other Acqn..}
      & \multicolumn{3}{v}{\bf AOL Specific}
      \\
      \hline\hline
      advertising.com	&	huffingtonpost.com	&	5min.com	&	aolcdn.com	&	advertising.aol.com	&	\\\hline
      adsonar.com	&	huffpo.net	&	tacoda.com	&	aol.com	&	atwola.com	&	\\\hline
      &	huffpost.com	&	{goviral-content.com}	&	srvntrk.com	&	blogsmithmedia.com	&	\\\hline
      &		&	mirabilis.com	&	mqcdn.com	&		&	\\\hline
      &		&	pictela.net	& mapquestapi.com	&		&	\\\hline
      \hline
    \end{tabular}
    \begin{tabular}{ g w g w w w }
      \multicolumn{6}{h}{\bf Conversant (former ValueClick)}
      \\
      \hline
      \multicolumn{1}{h}{\begin{minipage}{0.8in}{\bf Commission Junction}\end{minipage}}
      & \multicolumn{1}{v}{\bf Media Plex}
      & \multicolumn{1}{h}{\bf Other Acqn..}
      & \multicolumn{3}{v}{\bf Conversant Specific}
      \\
      \hline\hline
      yceml.net	&	mediaplex.com	&	dotomi.com	&	conversantmedia.com &  &	\\\hline
      ftjcfx.com	&	lduhtrp.net	&		&	apmebf.com &  &		\\\hline
      tqlkg.com	&		&		&	awltovhc.com     &  &	\\\hline
      qksrv.net	&		&		&	kdukvh.com &  &		\\\hline
      \hline
    \end{tabular}
    \begin{tabular}{ g w g g w w }
    
      \multicolumn{6}{h}{\bf Yahoo}
      
      \\
      \hline
      \multicolumn{1}{h}{\bf Flicker}
      & \multicolumn{1}{v}{\bf Yield Manager}
      & \multicolumn{2}{h}{\bf Other Acqn..}
      & \multicolumn{2}{v}{\bf Yahoo Specific}
      \\
      \hline\hline

      flickr.com	&	yieldmanager.com	&	bluelithium.com	&	overture.com	&	yahooapis.com	&	yimg.com	\\\hline
      staticflickr.com	&	yldmgrimg.net	&	maktoob.com	&	xtendmedia.com	&	yahoo.net	&	sstatic.net	\\\hline
      & & & &	&	zenfs.com	\\\hline

    \end{tabular}

    \caption{\label{tab:aqn}Hierarchical presentation of top four big companies, acquisitions and their third-party trackers.}
  \end{table*}
}

\section*{Appendix B}
\label{App:appendixb}

\begin{figure}
\includegraphics[trim=50 0 0 50, clip,width=0.37\textwidth]{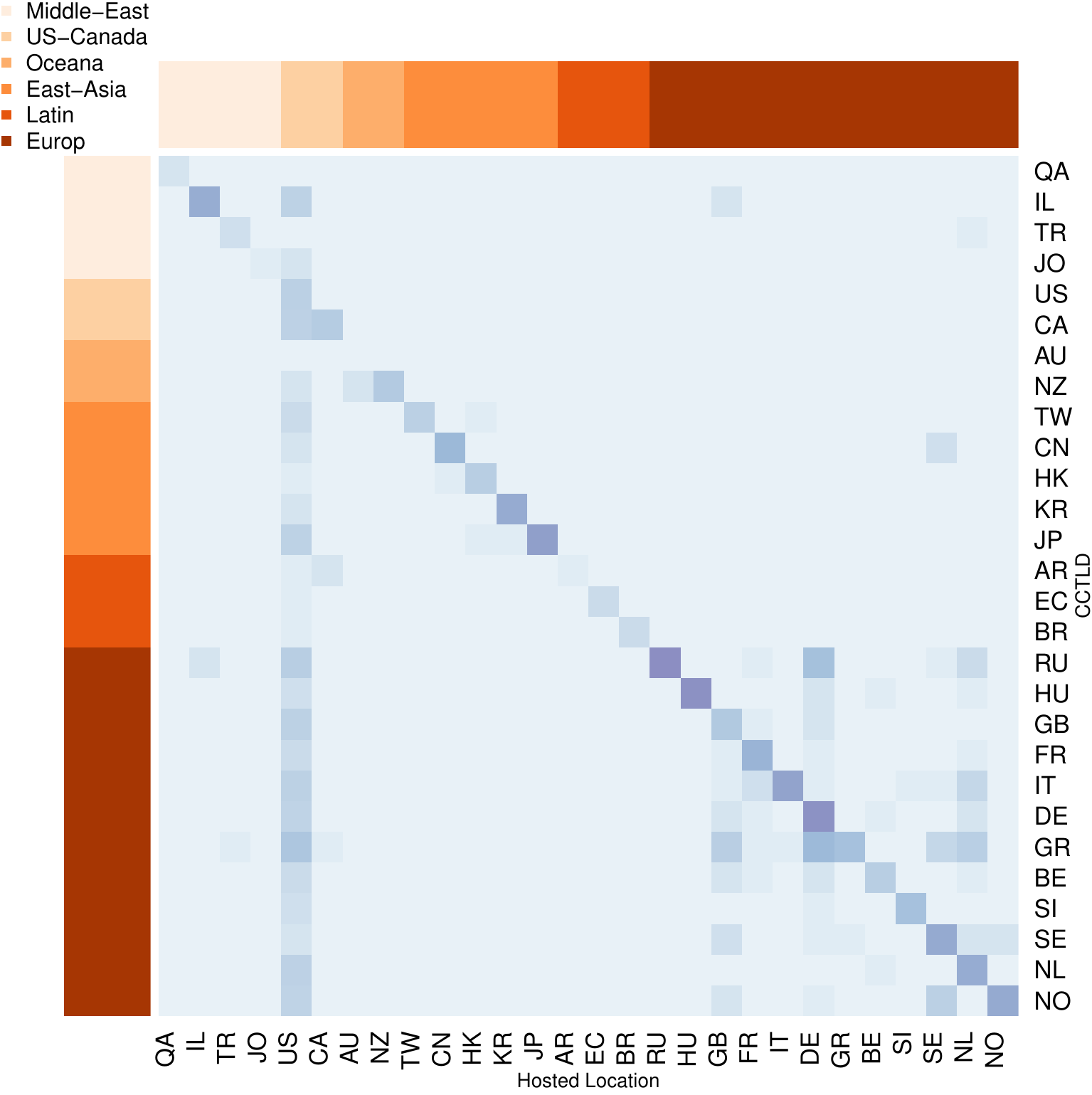}
\caption{Physical location of third-parties vs their CCTLD}
\label{map:host}

\end{figure}

The heat map in Figure~\ref{map:host} shows that the most of third-parties with CCTLDs are physically located in a country where their CCTDL point to. The $y$-axis represents the location of a third-party according to its country code, and the $x$-axis represents the physical location of the third-party. Darker colours indicate greater presence.

\end{document}